\documentclass[twocol]{ametsocV6.1}

\title{Tropical cyclone genesis potential using a ventilated potential intensity\\
{\color{red}NOT PUBLISHED. In peer review}  }

\authors{Daniel R. Chavas\aff{a}\correspondingauthor{Daniel R. Chavas, drchavas@gmail.com}
Suzana J. Camargo,\aff{b}, and Michael K. Tippett\aff{c}
}

\affiliation{\aff{a}{Purdue University, Department of Earth, Atmospheric, and Planetary Sciences, \\ West Lafayette, IN}\\
\aff{b}{Lamont-Doherty Earth Observatory, Columbia University, Palisades, NY} \\
\aff{c}{Department of Applied Physics and Applied Mathematics, Columbia University, New York, NY}
}

\abstract{Genesis potential indices (GPIs) are widely used to understand the climatology of tropical cyclones (TCs). However, the sign of projected future changes depends on how they incorporate environmental moisture. Recent theory combines potential intensity and mid-tropospheric moisture into a single quantity called the ventilated potential intensity, which removes this ambiguity. This work proposes a new GPI ($GPI_v$) that is proportional to the product of the ventilated potential intensity and the absolute vorticity raised to a power. This power is estimated to be approximately 5 by fitting observed tropical cyclone best-track and ECMWF Reanalysis v5 (ERA5) data. Fitting the model with separate exponents yields nearly identical values, indicating that their product likely constitutes a single joint parameter. Likewise, results are nearly identical for a Poisson model as for the power law. $GPI_v$ performs comparably well to existing indices in reproducing the climatological distribution of tropical cyclone genesis and its covariability with El Ni\~no-Southern Oscillation, while only requiring a single fitting exponent. When applied to Coupled Model Intercomparison Project Phase 6 (CMIP6) projections, $GPI_v$ predicts that environments globally will become gradually more favorable for TC genesis with warming, consistent with prior work based on the normalized entropy deficit, though significant changes emerge only at higher latitudes under relatively strong warming. $GPI_v$ helps resolve the debate over the treatment of the moisture term and its implication for changes in TC genesis favorability with warming, and its clearer physical interpretation may offer a step forward towards a theory for genesis across climate states.}

\begin{document}

\maketitle

%
%
%
\statement
Tropical cyclones cause significant human impacts globally, yet we currently don't understand what controls the number of storms that form each year. Tropical cyclone formation depends on fine-scale processes that our climate models cannot capture. Thus, it is common to use parameters from the background environment to represent regions favorable for cyclone formation. However, there are a variety of formulations because the link between environment and cyclone formation is complicated. This work proposes a new method that unifies a few common formulations, which helps resolve a divergence in current explanations of how tropical cyclone formation may change under climate change.

%
%
%

%




\section{Introduction}

Genesis potential indices (GPIs) are commonly used to explain the climatology of tropical cyclone activity on Earth from a small number of large-scale environmental parameters \citep{Camargo_Wing_2016,Sobel_etal_2021}. They have been employed to explain the observed mean-state climatology of tropical cyclogenesis \citep{Gray_1979,Camargo_etal_2007} as well as climatological variability induced by modes of climate variability on seasonal timescales, such as the El Ni\~no-Southern Oscillation \citep[ENSO;][]{Gray_1979,Camargo_Emanuel_Sobel_2007} and the Atlantic Meridional Mode \citep{Patricola_Saravanan_Chang_2014}, and subseasonal timescales, such as the Madden-Julian Oscillation \citep{Camargo_Wheeler_Sobel_2009,Klotzbach_Oliver_2015,Wang_Moon_2017}. GPIs have also been used to explain forced variations in tropical cyclone activity, such as due to volcanic eruptions \citep{Pausata_Camargo_2019}; paleoclimatic changes in orbital forcing \citep{Korty_Camargo_Galewsky_2012} and continental land surface properties \citep{Pausata_etal_2017}; and due to increasing greenhouse gases in the modern and future climate \citep{Camargo_2013,Korty_etal_2017,Emanuel_2021}.

Most GPIs contains two essential components, one dynamic and one thermodynamic, intended to capture the known physics required for the development and maintenance of a tropical cyclone. The dynamic component represents background rotation and is commonly represented by the lower-tropospheric (e.g., 850 hPa) absolute vorticity. Recent work has further argued for the inclusion of the low-level absolute vorticity gradient to represent the propensity to generate seed disturbances that precede genesis \citep{Tory_Ye_Dare_2018,Hsieh_eatl_2020}. The thermodynamic component represents the energetic potential of the atmosphere to support a tropical cyclone and is commonly defined by the maximum potential intensity \citep{Bister_Emanuel_2002}, the bulk deep-tropospheric (e.g., 850-200 hPa) wind shear, and a measure of the dryness of the mid-troposphere. The latter two terms together represent the import of environmental dry air into the storm inner core induced by the interaction of the vortex with background vertical wind shear \citep{Tang_Emanuel_2012}. These parameters are then combined together jointly \citep{Hoogewind_etal_2019}, in a simple multiplicative fashion allowing for different weighting exponents \citep{Emanuel_Nolan_2004,Emanuel_2021,Tory_Ye_Dare_2018} or in a Poisson framework \citep{Tippett_Camargo_Sobel_2011}. The dynamical GPI of \citet{Wang_Murakami_2020} deviates slightly from this structure by including an additional dynamical dependence on the meridional shear of the zonal wind specifically in the Southern Hemisphere and no explicit dependency on potential intensity or SST. Moreover, the mid-tropospheric vertical velocity has been proposed as an additional factor that measures seed activity \citep{Held_Zhao_2008,Hsieh_eatl_2020}. These variations in parameters and functional form for combining these parameters into an index reflects our lack of a proper theory for tropical cyclogenesis \citep{Sobel_etal_2021}, which fuels ongoing debate over the appropriate spatial and temporal scales for their use in explaining tropical cyclone variability \citep{Mei_etal_2019,Cavicchia_etal_2023}. While most GPIs were developed globally, some GPIs were especially designed for specific TC basins, with the argument that genesis processes are different across basins \citep[e.g.,][]{Bruyere_Holland_Towler2012,Meng_Garner_2023}. Recently, machine learning methods have started to be used to develop alternative GPIs \citep{Fu_Chang_Liu2023}. 

Ultimately, many GPIs share the same dynamic component (low-level absolute vorticity) and share a common structure for the thermodynamic component: they start from the potential intensity and then modify it based on the known effects of wind shear (negative) and environmental moisture (positive). Of these two factors, the representation of environmental mid-level moisture has caused the greatest consternation, with choices spanning relative humidity, saturation deficit, and normalized entropy deficit as detailed in Section \ref{sec:methods} below. This choice has significant consequences in projections of future TC genesis with warming: relative humidity and normalized entropy deficit yield an increase and saturation deficit yields a decrease \citep{Camargo_etal_2014,Lee_etal_2020,Lee_etal_2023,Emanuel_2021}. This degree of epistemic uncertainty translates to very large uncertainty in how tropical cyclone activity will change under future warming, including a lack of consensus on the sign of the change in storm counts \citep{Knutson_etal_2020,Camargo_etal_2023}.

Recently, \cite{Komacek_Chavas_Abbot_2020} derived an analytic expression for the ventilated potential intensity in the context of understanding the potential for tropical cyclone activity on tidally-locked exoplanets. Their expression is derived directly from the foundational theory for how dry air ventilation acts to reduce the traditional potential intensity below its nominal (ventilation-free) value \citep{Tang_Emanuel_2010}. The effects of ventilation are captured by a single quantity known as the Ventilation Index, which depends specifically on the normalized entropy deficit and is known to be a very useful predictor for both genesis and intensification rate in nature \citep{Tang_Emanuel_2012}. The ventilation index motivates the GPI of \cite{Emanuel_2010}, employed in \cite{Emanuel_2021}, in which the normalized entropy deficit, vertical wind shear, and potential intensity are taken as separate predictors. An alternative and more direct approach, though, would be to use the theoretical prediction for the ventilated potential intensity itself, thereby integrating all three thermodynamic terms together into a single parameter that is fully rooted in a successful theory for tropical cyclone energetics. This approach has the opportunity to greatly simplify existing genesis parameters and reduce the degrees of freedom of our genesis parameters and their wide range of outcomes. Moreover, a simpler GPI may yield insights into the genesis process that can better link our GPIs to a broader theory for genesis. Indeed, \cite{Komacek_Chavas_Abbot_2020} defined TC genesis favorability by combining the ventilated potential intensity with the 850-hPa absolute vorticity, and \cite{Garcia_etal_2024} found that this combination can successfully predict genesis regions from explicitly-simulated TCs in idealized high-resolution exoplanet simulations. This approach has yet to be applied to Earth, which is the focus of our work here.

This work examines the use of a ventilated potential intensity in GPI formulations. Our aim is to demonstrate a novel and simpler approach to representing the thermodynamic component of genesis potential relative to existing versions and explore its climatological applications. Specifically, we aim to answer the following questions:
\begin{enumerate}
    \item What is the climatological distribution of the ventilated potential intensity?
    \item To what extent can a GPI be simplified when using the ventilated potential intensity?
    \item How well does this simplified GPI capture the climatology of tropical cyclogenesis, including its seasonal cycle and ENSO? 
    \item How do the results compare to existing GPI formulations with comparable formulations, including changes in future projections?
\end{enumerate}
We answer these questions via historical tropical cyclone genesis data and environmental data from reanalysis and future climate change projection simulations. For the final question regarding climate change, our hypothesis is that a GPI based on the ventilated potential intensity should predict an increase in TC genesis with warming as was found in \cite{Emanuel_2021} given their shared theoretical basis.

Section 2 revisits the basic theory of the ventilated potential intensity.  Section 3 describes the data and methods. Section 4 presents the results. Section 5 summarizes key results and discusses avenues of future work.

\section{Theory}

\subsection{Ventilated Potential Intensity}

\begin{figure*}[t]
\centering
 \noindent\includegraphics[width=0.6\textwidth]{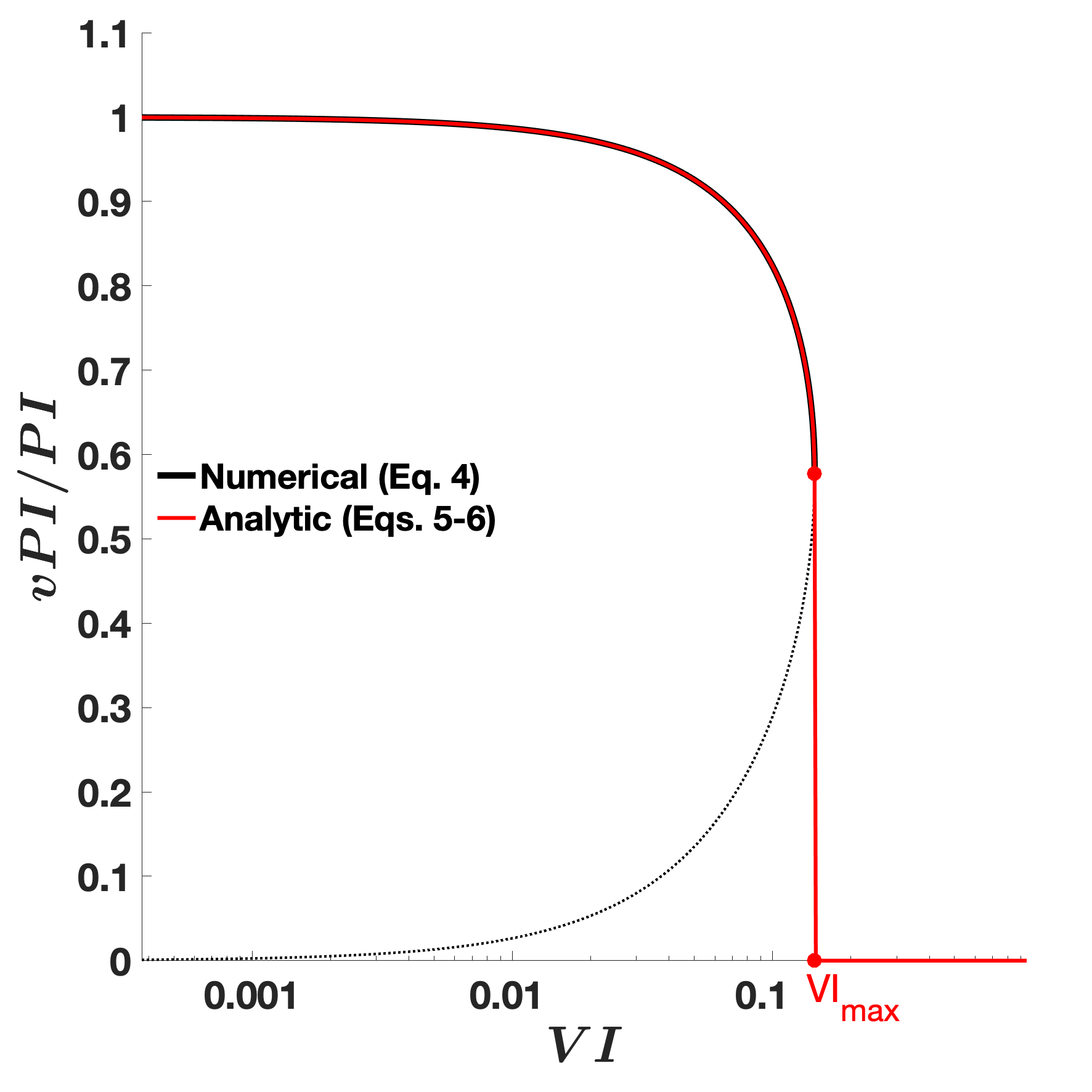}\\
 \caption{Ventilated potential intensity, $vPI$, as a fraction of the standard (ventilation-free) potential intensity, $PI$. Analytic solution (red) given by Eq. \eqref{eq:PIvent1}-\eqref{eq:PIvent2}. Numerical solution to Eq. \eqref{eq:PIvent_equil} is shown in black, with stable equilibrium in solid and unstable equilibrium in dashed. $VI_{max}$ is the maximum VI value that yields a non-zero $vPI$, which is set here to $VI_{max} = 0.145$ following \cite{Hoogewind_etal_2019}.}\label{fig:PIventtheory}
\end{figure*}

We begin with a brief re-derivation of the ventilated potential intensity ($vPI$) initially developed in \cite{Komacek_Chavas_Abbot_2020}, whose final form was simplified in \cite{Garcia_etal_2024}. The solution is visualized in Figure \ref{fig:PIventtheory}.

The potential intensity ($PI$) is the upper bound on the maximum wind speed that can be attained in the hurricane eyewall for a given thermodynamic environment \citep{Emanuel_1986}. Its basis is in the energetics of a differential heat engine across the eyewall, where the system powers its circulation against surface friction by adding entropy at the warm sea surface and losing entropy in the cold outflow \citep{Rousseaurizzi_Emanuel_2019}. This quantity is given by
\begin{equation}
    PI^2 = \frac{C_k}{C_d}\frac{T_s-T_o}{T_o}\left(k^*_s - k\right)
\end{equation}
where $T_s$ is the surface temperature, $T_o$ is the outflow temperature, $\frac{C_k}{C_d}$ is the ratio of the exchange coefficients of enthalpy and momentum, and $k^*_s - k$ is the difference between the convective available potential energy (CAPE) of a hypothetical saturated air parcel at sea level temperature and pressure and a boundary layer air parcel, both calculated at the radius of maximum wind. Estimation of the outflow temperature and the pressure dependence of the enthalpy at the radius of maximum wind both require numerical integration, which is commonly calculated using the algorithm of \cite{Bister_Emanuel_2002}. We use this algorithm here, too.

The ventilation index (VI) of \cite{Tang_Emanuel_2012} represents the potential import of environmental dry air into the storm inner core by environmental vertical wind shear. This quantity is given by
\begin{equation}
    VI = \frac{V_s\chi_m}{PI}
\end{equation}
where $V_s$ is the magnitude of the vector wind difference between 850 hPa and 200 hPa and $\chi_m$ is the mid-troposphere entropy deficit calculated at 600 hPa, following \cite{Hoogewind_etal_2019}. The entropy deficit is given by
\begin{equation}
    \chi_m = \frac{s^*_m - s_{m,env}}{s^*_{SST}-s_b}
\end{equation}
where $s^*_m - s_{m,env}$ is the difference between the saturation entropy in the eyewall and the environmental entropy, and $s^*_{SST}-s_b$ is the difference between the saturation entropy at the sea surface temperature and the entropy of the boundary layer air parcel.

Ventilation acts as ``anti-fuel" that weakens the hurricane heat engine and thus reduces the true potential intensity from its ventilation-free value \citep{Tang_Emanuel_2012}. The ventilated potential intensity ($vPI$) represents a generalized form of $PI$ that includes the effect of ventilation. This quantity is derived from the equilibrium solution for normalized intensity in the presence of ventilation \citep{Chavas_2017}
\begin{equation}\label{eq:PIvent_equil}
    \widetilde{vPI}^3 - \widetilde{vPI} + \frac{2}{3\sqrt{3}}\frac{VI}{VI_{max}} = 0    
\end{equation}
where $\widetilde{vPI} = \frac{vPI}{PI}$ is the $vPI$ as a fraction of the ventilation-free $PI$, and $VI_{max}$ is a threshold value of VI representing the maximum value of VI where $vPI>0$. The third term is the referred to as the normalized ventilation \citep{Tang_Emanuel_2012}. When there is zero ventilation ($VI=0$), the solution to Eq. \ref{eq:PIvent_equil} is $vPI = PI$, i.e. there is no reduction in the potential intensity from its standard value. As $VI$ increases above zero, $vPI$ is more strongly reduced from the standard $PI$. For $VI = VI_{max}$, $vPI$ is equal to 57.74\% of the $PI$. Above this threshold ($VI > VI_{max}$), the solution drops immediately to $vPI = 0$.

\cite{Komacek_Chavas_Abbot_2020} derived an analytic solution to Eq. \ref{eq:PIvent_equil}, which can be written in a simpler form following some mathematical reduction \citep{Garcia_etal_2024} as:
\begin{equation}\label{eq:PIvent1}
    vPI = \left(x+\frac{1}{3x}\right)PI
\end{equation}
where
\begin{equation}\label{eq:PIvent2}
    x = \frac{1}{\sqrt{3}}\left[\sqrt{\left(\left(\frac{VI}{VI_{max}}\right)^2-1\right)}-\left(\frac{VI}{VI_{max}}\right)\right]^\frac{1}{3}
\end{equation}
This solution can have both real and imaginary components; only the real component is physical and is used as the solution. This analytic solution is shown in Figure \ref{fig:PIventtheory}.

As can be seen in Figure \ref{fig:PIventtheory}, a major conceptual benefit of $vPI$ is that this quantity cuts off sharply to zero at the threshold value of VI ($VI_{max}$). This property allows for sharper cutoffs in genesis regions, which may be valuable given the propensity for existing GPIs to be too gradual in the climatological transition from favorable to unfavorable regions.

Similar to the standard potential intensity, $vPI$ may be readily calculated from a vertical profile of temperature and moisture and a sea surface temperature and pressure. This quantity will be calculated at every gridpoint in reanalysis fields and climate model simulations as described below.

\subsection{GPI with ventilated potential intensity ($GPI_v$)}

The thermodynamic parameter for our GPI is given by $vPI$ given by Eqs. 5--6. For the dynamical parameter, we follow existing GPIs and use the ``clipped'' 850-hPa absolute vorticity given by
\begin{equation}
    \eta_c = min[3.7 \times 10^{-5} , f + \zeta_{850}]
\end{equation}
where $f = 2\Omega sin(\frac{\pi}{180}\phi)$ is the Coriolis parameter, with $\phi$ the latitude in degrees and $\Omega = 7.292*10^{-5} \: s^{-1}$ the Earth's rotation rate, and $\zeta_{850} = \hat{k} \cdot \nabla \times \vec{\textbf{u}}_{850}$ is the relative vorticity calculated at 850 hPa. We then set an upper bound on its value of $3.7\times 10^{-5} \: s^{-1}$ following \cite{Tippett_Camargo_Sobel_2011} and consistent with the existence of a latitude of maximum genesis rate in aquaplanet simulations of \cite{Chavas_Reed_2019}. The existence of this upper bound indicates a threshold of ``sufficient'' background rotation above which there is no additional benefit to promoting genesis.




%
We test two standard formulations for combining the thermodynamic and dynamic quantities to construct a GPI.  The first has a power law form $\left( c * vPI * \eta_c \right)^\alpha \cos \phi$, where $\phi$ is latitude in radians; the $\cos \phi$ term accounts for the reduction in surface area with latitude (i.e., a latitude-longitude grid box is smaller at higher latitudes so is intrinsically less likely to contain pointwise genesis events). The exponent $\alpha$ and normalizing factor $c$ are found by constrained least-squares with the constraint that global annual mean match the observed value of 84.625 (1981--2020). The final result is
\begin{equation}\label{eq:GPIvent_pl}
    GPI_v = \left( 135.5 * vPI * \eta_c \right)^{4.90} \cos \phi\,,
\end{equation}
When fitting the model with separate exponents for the two components, the best exponents are $4.85$ for $vPI$ and $4.97$ for $\eta_c$. These values are very similar, which indicates that the two parameters may simply be combined together as we have done in Eq. \eqref{eq:GPIvent_pl} above. Additionally, when fitting the model without the least-squares constraint for the model to match the global-mean number (i.e., $c$ is a 2nd free parameter in the fit), the exponent is $5.07$, which is very similar to the one-parameter constrained model above; hence the latter is preferable and is used here.

The second GPI formulation has the form of a Poisson regression given by $\exp\left(\beta_0 + \beta_1 * vPI * \eta_c \right) \cos \phi$. The parameters $\beta_0$ and $\beta_1$ are maximum likelihood estimates with $\cos \phi$ entering the Poisson regression as an offset. The final result is
\begin{equation}\label{eq:GPIvent_pois}
GPI_{v,Pois} = \exp\left(-10.13 + 0.0188 * vPI * \eta_c \right) \cos \phi\,.
\end{equation}
A convenient property of the Poisson regression is that it matches the global mean by construction. Squared error was reduced by less than 0.1\% when allowing separate exponents for $vPI$ and $\eta_c$ in the power law or separate regression coefficients in the Poisson regression. This outcome again indicates that the two quantities can indeed be combined into a single predictor.

\begin{figure*}[t]
 \begin{center}
 \noindent\includegraphics[width=0.8\textwidth]{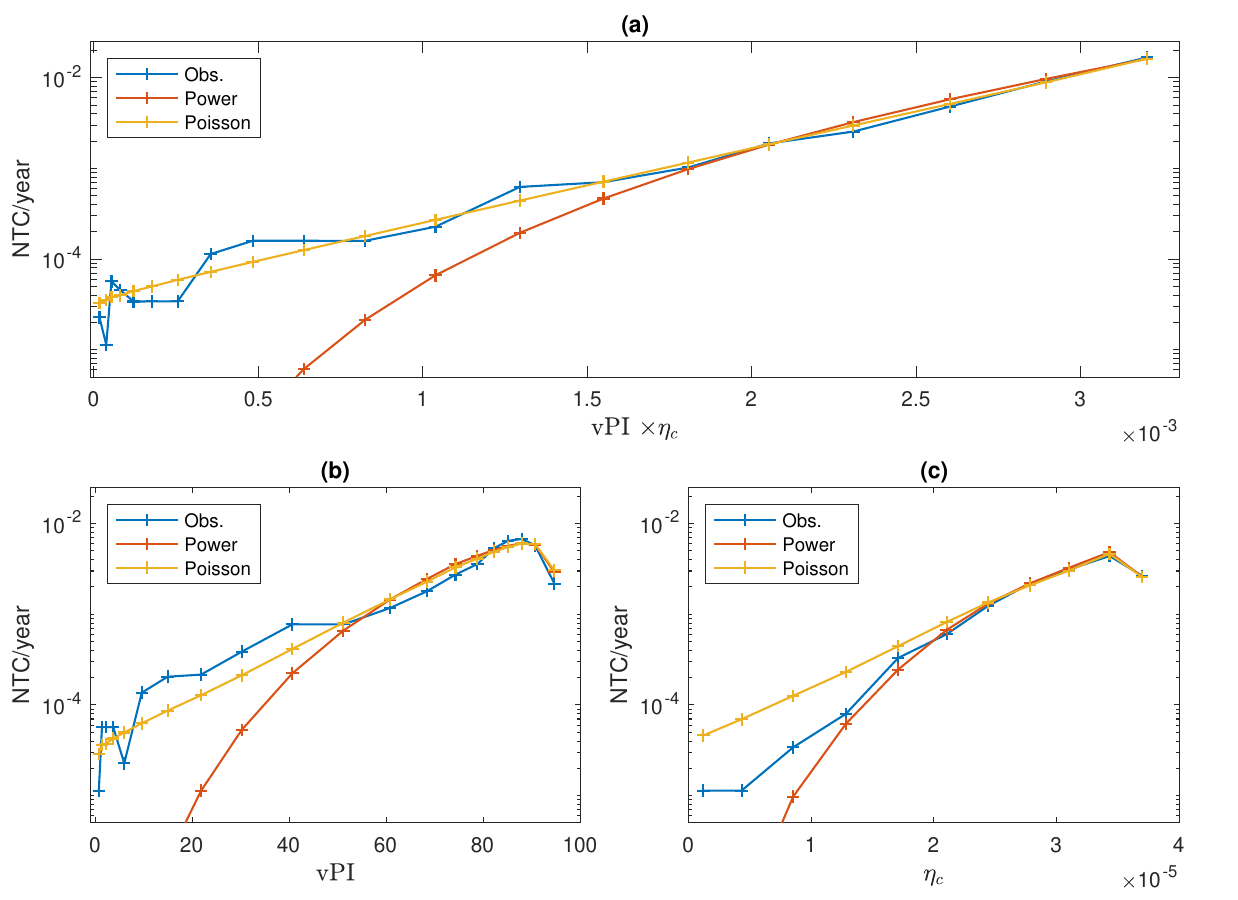}\\
 \end{center}
 \caption{Dependence of observations (blue), $GPI_v$ (red), and $\text{GPI}_{\text{vent,Pois}}$ on (a) $vPI * \eta_c$, (b) $vPI$, and (c) $\eta_c$. Plots are displayed as log-linear, such that an exponential (Poisson) dependence will be linear. Abscissa values are separated into 20 bins (equal numbers of values).
 }\label{fig:GPIv_pl_pois}
\end{figure*}

We choose the power-law form, as the behavior of the power-law and Poisson formulations are very similar. Fig.\ \ref{fig:GPIv_pl_pois}a compares the observed and predicted dependence of the number of TCs per year on $vPI * \eta_c$. For the range of values greater than $1.5\times 10^{-3}$, which represents the vast majority of all TCs (note the log-scale on the y-axis; TC frequency at this threshold is approximately 1/20 of the peak), the two models are nearly identical. For smaller values they diverge and Poisson is the better model, as the observed dependence in the log-linear plot of Fig.\ \ref{fig:GPIv_pl_pois}a is nearly linear, consistent with the exponential dependence of the Poisson model that is also linear (a power law model is linear in a log-log space). However, this departure at small values of $vPI * \eta_c$ is associated with regions/times on Earth that have very few TCs at all: small values of $vPI\le40 \: ms^{-1}$ (Fig.\ \ref{fig:GPIv_pl_pois}b), corresponding to regions on Earth that cannot support TCs at a reasonable intensity, and $\eta_c < 1.25$ (Fig.\ \ref{fig:GPIv_pl_pois}c), corresponding to very low equivalent latitudes of $<5^\circ$ where storm formation is very rare \citep{Chavas_Reed_2019}.\footnote{Mathematically, power law and exponential functions are very similar so long as the range of their shared predictor is not too large. A power law is given by $\log(y)=\log(a_1) + a_2\log(x')$, while an exponential is given by $\log(y) = \log(b_1)+b_2x'$, where $x'=x-x_0$ is the departure from the $y$-weighted mean value $x_0$ to which both models are fit. The difference is strictly $\log(x')$ vs $x'$ on the right hand sides. Substituting in the Taylor series approximation $\log(x') \approx (x'-1) - \frac{1}{2}(x'-1)^2$, where higher order terms have been dropped, and rearranging yields $\log(y)=(\log(a_1) - a_2) + a_2x' - \frac{a_2}{2}(x'-1)^2$. This equation now has the same form as an exponential function, with the differences scaling as $\frac{a_2}{2}(x'-1)^2$. Hence, the models will be very similar in the vicinity of $x_0$ where the bulk of the data resides and will differ significantly only when moving sufficient far from $x_0$.} Hence, for practical purposes the power law formulation (Eq. \ref{eq:GPIvent_pl}) and Poisson formulation (Eq. \ref{eq:GPIvent_pois}) are equivalent. This close equivalence between the two forms has not to our knowledge been addressed in past literature despite both forms being consistently used in different GPI formulations \citep{Emanuel_Nolan_2004,Tippett_Camargo_Sobel_2011}. Their mathematical similarity over the range of predictor values relevant to the vast majority of TCs on Earth may explain why both formulations perform similarly in practice.

Given their similar practical performance, for the remainder of our study we utilize the power law model of Eq. \eqref{eq:GPIvent_pl}. Indeed, all of our results below are nearly indistinguishable if the Poisson formulation is used instead. Additionally, the power law formulation has important intuitive and physical appeal. First and foremost, the quantity $vPI * \eta_c$ has units of acceleration ($[m \: s^{-2}]$), which might be interpreted as analogous to an intensification rate. Second, past work has argued that vertical velocity $w \sim PI$ \citep{Khairoutdinov_Emanuel_2013} in a tropical cyclone, in which case this quantity behaves like $w\eta$, i.e. a vortex stretching, which is the lone vorticity source term in the vorticity equation proposed as a genesis framework by \cite{Hsieh_eatl_2020}. This interpretation also aligns at least qualitatively with the importance of the overall updraft mass flux in setting regions of significant convective activity and precursor seed disturbances \citep{Held_Zhao_2008,Hsieh_eatl_2020}. The exponent represents a strong sensitivity of genesis to this quantity, but its magnitude lacks any obvious physical explanation. Nonetheless, this GPI formulation is simpler and may take an important step closer to linking the GPI to a physical theory of genesis; further discussion is provided in Section 5 below.

\section{Data and Methods}\label{sec:methods}

Global historical 6-hourly tropical cyclone genesis data are taken from the IBTrACS data archive \citep{Knapp_etal_2010}. Environmental parameters are calculated from monthly mean ERA5 reanalysis fields \citep{Hersbach_etal_2020}. The horizontal grid spacing of ERA5 is 0.25 degree and all fields were interpolated to a 2 degree horizontal spacing. For both datasets, we analyze the period 1981–2020.

For historical climate model simulation data, we use monthly surface and pressure-level data for 1981–2014 from 45 CMIP6 model historical simulations \citep{Eyring_etal_2016}. For future climate simulation projections, we use data from the ssp245, ssp370, and ssp585 scenarios and analyze each for the periods 2021--2040 (near-term), 2041--2060 (medium-term), and 2071--2100 (long-term). All model data were interpolated to a common 2 degree horizontal spacing. For each model and scenario, all ensembles which had all required output fields for all variables were considered. The ensemble mean of each model was calculated using all the ensembles available before calculating the multi-model mean for each scenario. Note that the number of models available differed across future scenarios, as well as the number of ensembles. When the differences between future and current scenarios are calculated, the models in the historical scenario matched those in each of the future scenarios. Table~\ref{tab:ModelsNumbEns} lists the models and the number of ensembles in each scenario as well as the papers that describe each of these models.

\begin{table*}
\begin{center}
\begin{tabular}{|c|c|c|c|c|c|l|}
\hline
Number & Model & Historical & ssp245 & ssp370 & ssp585 & References\\ \hline 
1 & ACCESS-CM2 & 3 & 3 & 3 & 5 & \citet{biConfigurationSpinupACCESSCM22020}\\
2 & ACCESS-ESM1-5 & 20 & 11 & 10 & 40 & \citet{ziehnAustralianEarthSystem2020}\\
3 & AWI-CM-1-1-MR & 5 & 1 & 5 & 1  & \citet{semmlerSimulationsCMIP6AWI2020}\\
4 & BCC-CSM2-MR & 3 & 1 & 1 & 1 & \citet{wuBeijingClimateCenter2019}\\
5 & CAMS-CSM1-0 & 3 & 2 & 2 & 2 & \citet{rongCAMSClimateSystem2018,rongCMIP6HistoricalSimulation2021}\\
6 & CanESM5 & 30 & 50 & 50 & 50 & \citet{swartCanadianEarthSystem2019}\\
7 & CanESM5-CanOE & 3 & 3 & 3 & 3 & \citet{swartCanadianEarthSystem2019}\\
8 & CAS-ESM2-0 & 4 & 2 & 1 & 2 & \citet{zhangDescriptionClimateSimulation2020}\\
9 & CESM2 & 11 & 6 & 6 & 5 & \citet{danabasogluCommunityEarthSystem2020}\\
10 & CESM2-WACCM & 3 & 5 & 3 & 5 & \citet{danabasogluCommunityEarthSystem2020}\\
11 & CIESM & 3 & 1 & --- & 1 & \citet{linCommunityIntegratedEarth2020}\\
12 & CMCC-CM2-SR5 & 1 & 1 & 1 & 1 & \citet{lovatoCMIP6SimulationsCMCC2022}\\
13 & CMCC-ESM2 & 1 & 1 & 1 & 1 & \citet{lovatoCMIP6SimulationsCMCC2022}\\
14 & CNRM-CM6-1 & 28 & 10 & 6 & 6 & \citet{voldoireEvaluationCMIP6DECK2019}\\
15 & CNRM-CM6-1-HR & 1 & 1 & 1 & 1  & \citet{voldoireEvaluationCMIP6DECK2019}\\
16 & CNRM-ESM2-1 & 9 & 10 & 5 & 5 & \citet{seferianEvaluationCNRMEarth2019}\\
17 & E3SM-1-1 & 1 & --- & --- & 1 & \citet{golazDOEE3SMCoupled2019}\\
18 & EC-Earth3 & 18 & 22 & 7 & 8 & \citet{doscherECEarth3EarthSystem2022}\\
19 & EC-Earth3-CC & 1 & 1 & --- & 1 & \citet{doscherECEarth3EarthSystem2022}\\
20 & EC-Earth3-Veg & 9 & 6 & 6 & 8 & \citet{doscherECEarth3EarthSystem2022}\\
21 & EC-Earth3-Veg-LR & 3 & 3 & 3 & 3 & \citet{doscherECEarth3EarthSystem2022}\\
22 & FGOALS-f3-L & 3 & 1 & 1 & 1 & \citet{heCASFGOALSf3LModel2019}\\
23 & FGOALS-g3 & 6 & 4 & 5 & 4 & \citet{liFlexibleGlobalOceanAtmosphereLand2020}\\
24 & FIO-ESM-2-0 & 3 & 3 & --- & 3 & \citet{baoFIOESMVersionModel2020}\\
25 & GFDL-CM4 & 1 & 1 & --- & 1 & \citet{heldStructurePerformanceGFDL2019}\\
26 & GFDL-ESM4 & 4 & 3 & 1 & 1 & \citet{dunneGFDLEarthSystem2020}\\
27 & GISS-E2-1-G & 47 & 20 & 18 & 11 & \citet{kelleyGISSE2ConfigurationsClimatology2020,millerCMIP6HistoricalSimulations2021}\\
28 & GISS-E2-1-H & 25 & --- & --- & 5 & \citet{kelleyGISSE2ConfigurationsClimatology2020,millerCMIP6HistoricalSimulations2021}\\
29 & HadGEM3-GC31-LL & 5 & 4 & --- & 4 & \citet{kuhlbrodtLowResolutionVersionHadGEM32018,andrewsForcingsFeedbacksClimate2019}\\
30 & HadGEM3-GC31-MM & 2 & ---  &  ---  & 4 & \citet{kuhlbrodtLowResolutionVersionHadGEM32018,andrewsForcingsFeedbacksClimate2019}\\
31 & INM-CM4-8 & 1 & 1 & 1 & 1 & \citet{volodinSimulatingPresentdayClimate2010a}\\
32 & INM-CM5-0 & 10 & 1 & 5 & 1 & \citet{volodinSimulationPresentdayClimate2017}\\
33 & IPSL-CM6A-LR & 30 & 8 & 11 & 6 & \citet{boucherPresentationEvaluationIPSLCM6ALR2020}\\
34 & KACE-1-0-G & 3 & 3 & 3 & 3 & \citet{leeEvaluationKoreaMeteorological2020}\\
35 & KIOST-ESM2-0 & 1 & --- & --- & 1 & \citet{pakKoreaInstituteOcean2021}\\
36 & MIROC6 & 50 & 3 & 3 & 50 & \citet{tatebeDescriptionBasicEvaluation2019}\\
37 & MIROC-ES2L & 30 & 30 & 10 & 10 & \citet{hajimaDevelopmentMIROCES2LEarth2020}\\
38 & MPI-ESM1-2-HR & 10 & 2 & 10 & 1 & \citet{mullerHigherresolutionVersionMax2018,gutjahrMaxPlanckInstitute2019}\\
39 & MPI-ESM1-2-LR & 10 & 10 & 10 & 10 & \citet{mauritsenDevelopmentsMPIMEarth2019a}\\
40 & MRI-ESM2-0 & 6 & 2 & 5 & 4 & \citet{yukimotoMeteorologicalResearchInstitute2019}\\
41 & NESM3 & 5 & 2 & --- & 2 & \citet{caoNUISTEarthSystem2018}\\
42 & NorESM2-LM & 3 & 3 & 3 & 1 & \citet{selandOverviewNorwegianEarth2020}\\
43 & NorESM2-MM & 2 & 2 & 1 & 1 & \citet{selandOverviewNorwegianEarth2020}\\
44 & TaiESM1 & 2 & 1 & 1 & 1 & \citet{wangPerformanceTaiwanEarth2021}\\
45 & UKESM1-0-LL & 18 & 5 & 10 & 5 & \citet{sellarUKESM1DescriptionEvaluation2019}\\
\hline
\end{tabular}
\vspace{0.3cm}
\caption{CMIP6 models, number of ensembles per scenario and references}.
\label{tab:ModelsNumbEns}
\end{center}
\end{table*}

We first calculate historical tropical cyclogenesis rates using the IBTrACS dataset across all major basins globally as our observational baseline for the period 1981--2020. The Eastern Pacific basin is defined as all storms east of 180 W. We calculate our new GPI as well as existing commonly-used GPIs from environmental data in ERA5 reanalysis for comparison against one another and against observed genesis rates. We then calculate our new GPI in CMIP6 historical and future simulations to examine future changes under global warming. Note that the CMIP6 historical simulations in the CMIP6 models end in 2014. Therefore, when calculating the historical climatology we used the closest full decades to the observational climatology 1981--2010. When considering changes between present and future simulations we considered the last 30 years of the 20th Century (1971-2000) and the last 30 years of the 21st Century (2071--2100).    

Additionally, we compare our new GPI against four common existing GPIs. The first two are 
\begin{equation}
    \text{GPI}_{E04} = \left|10^5 \eta \right|^{3/2}\left(\frac{h_{600}}{50}\right)^3\left(\frac{PI}{70}\right)^3\left(1+0.1V_s\right)^{-2},
\end{equation}
where $h_{600}$ is the relative humidity at 600hPa in percentage, from \cite{Emanuel_Nolan_2004}, and
\small
\begin{equation}
    \text{GPI}_{E10} = \left|\eta \right|^3\chi_m^{-\frac{4}{3}}\max\left[\left(PI - 35 \: ms^{-1}\right),0\right]^2\left(25 \: ms^{-1} + V_s\right)^{-4}
\end{equation}
\normalsize
from \cite{Emanuel_2010}.
The second two are distinct versions of the Tropical Cyclone Genesis Index \cite[TCGI; ][]{Tippett_Camargo_Sobel_2011} using the formulations provided in \cite{Camargo_etal_2014}, with coefficients obtained from a Poisson regression of the ERA5 environmental fields. Note that different reanalyses products lead to slightly different coefficients and values for TCGI, as shown in \citet{Dirkes_etal_2023}. The first takes column relative humidity as the free-tropospheric moisture variable
\begin{equation}
    \text{TCGI}\mbox{--}CRH = \exp\left(b + b_\eta \eta_c + b_H H + b_{PI}PI + b_V V_s\right)
\end{equation}
where $H$ is the ratio of the column-integrated water vapor $W$ and its saturated value $W^*$, $H = W/W^*$. In the case of TCGI-CRH the coefficients are $b=-24.132$, $b_{\eta}=2.512$, $b_{H}=0.077$, and $b_{V}=-0.12$.The second takes column saturation deficit as the free-tropospheric moisture variable
\begin{equation}
    \text{TCGI}\mbox{--}SD = \exp\left(b + b_\eta \eta_c + b_{SD} SD + b_{PI}PI + b_V V_s\right)
\end{equation}
where $SD$ is the difference $SD = W-W^*$ (defined negative, following \citet{Bretherton_etal_2004}). The coefficients for TCGI-SD are $b=-18.356$, $b_{\eta}=2.483$, $b_{SD}=0.073$, $b_{PI}=0.08$ and $b_{V}=-0.135$. Both are used to examine future changes in TC activity \citep{Lee_etal_2020,Lee_etal_2022,Lee_etal_2023}. TCGI integrates to the global TC count (84.625) by design. To facilitate comparison, we normalize the two Emanuel indices to match this value as well. 

\section{Results}

We begin first with the climatology of our new genesis potential index $GPI_{v}$ and its components, including the ventilated potential intensity $vPI$. We then compare $GPI_v$ with existing GPIs. Finally, we examine how $GPI_v$ changes in the future in CMIP6 future climate projection simulations.

\subsection{Climatology of $vPI$, $GPI_v$}

\subsubsection{Annual climatology}


\begin{figure*}[t]
 \begin{center}
 \noindent\includegraphics[width=0.8\textwidth]{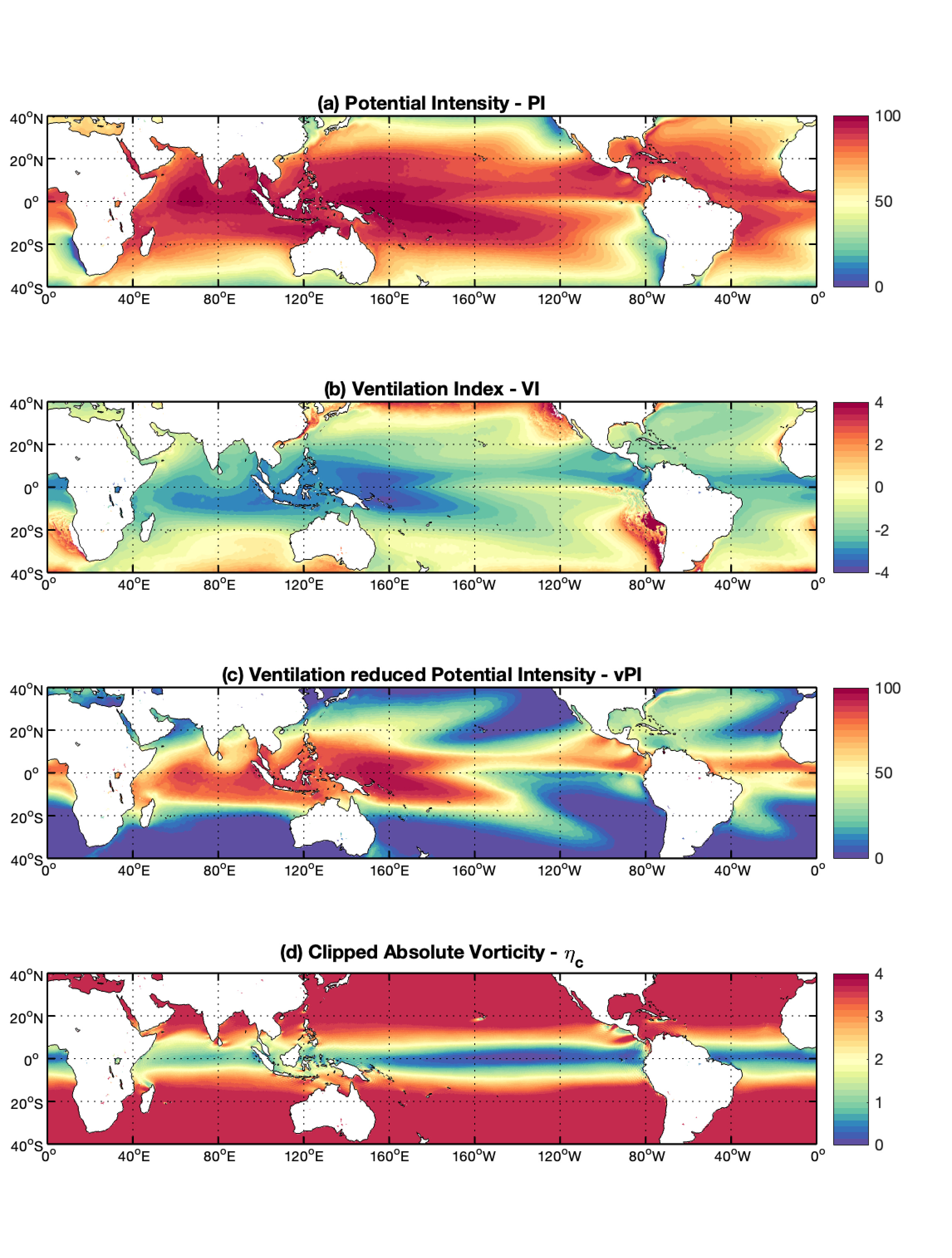}\\
 \end{center}
 \caption{Maps of annual-mean climatology of (a) Potential intensity ($PI$); (b) ventilation index (VI); (c) ventilated potential intensity ($vPI$); and (d) 850-hPa clipped absolute vorticity $\eta_c$. Data from ERA5 1981--2020.}\label{fig:vPIPIabsvort}
\end{figure*}

The annual-mean climatology of the components of $GPI_v$ are shown in Figure \ref{fig:vPIPIabsvort}. The traditional $PI$ (Figure \ref{fig:vPIPIabsvort}a) takes higher values broadly over the equatorial and western portions of the global tropical oceans and decreases gradually moving poleward and eastward within each basin. The transition from $PI$ to the ventilated potential intensity $vPI$ is mediated by the deleterious effects of the ventilation index (Figure \ref{fig:vPIPIabsvort}b; Eq. \ref{eq:PIvent1}). The regions of lowest ventilation (i.e., smaller values of VI), which are favorable for TCs, are also found in the same general regions as where potential intensity is high. The resulting distribution of $vPI$ (Figure \ref{fig:vPIPIabsvort}c) is a qualitatively similar pattern to $PI$ but with a much sharper transition from high values, primarily over the western North and South Pacific Ocean and the near-equatorial Indian Ocean and North Atlantic Ocean, to very low values moving poleward and eastward from those regions. In the regions of highest $PI$ over the equatorial Western Pacific Ocean, the ventilation index is very small and hence there is very little reduction in $vPI$ relative to $PI$. In contrast, moving poleward and eastward within each basin, particularly the North and South Atlantic and Pacific basins, the ventilation index increases substantially and thus there is a sharp reduction in $vPI$, decreasing rapidly towards zero as the ventilation index approaches its maximum value that allows a positive $vPI$. Hence, $vPI$ yields a much more refined geographic region capable of supporting tropical cyclones than $PI$.

The distribution of clipped absolute vorticity (Figure \ref{fig:vPIPIabsvort}d) is largely zonal in structure with values monotonically increasing moving poleward, reflecting the dominant contribution of $f$, up to the cut-off value above which it is held constant. The lone notable large-scale deviation is over the Indian Ocean and Maritime Continent in the Western Pacific, where the absolute vorticity is relatively large owing to a significant time-mean relative vorticity contribution from a quasi-stationary cyclonic circulation. This equatorial region is characterized by relatively high values of both absolute vorticity and ventilated potential intensity, and indeed it coincides with tropical cyclone formation closer to the equator than anywhere else in the world \citep{Lu_Ge_Peng_2021,Chavas_Reed_2019}.

\begin{figure*}[t]
 \begin{center}
\noindent\includegraphics[width=0.9\textwidth]{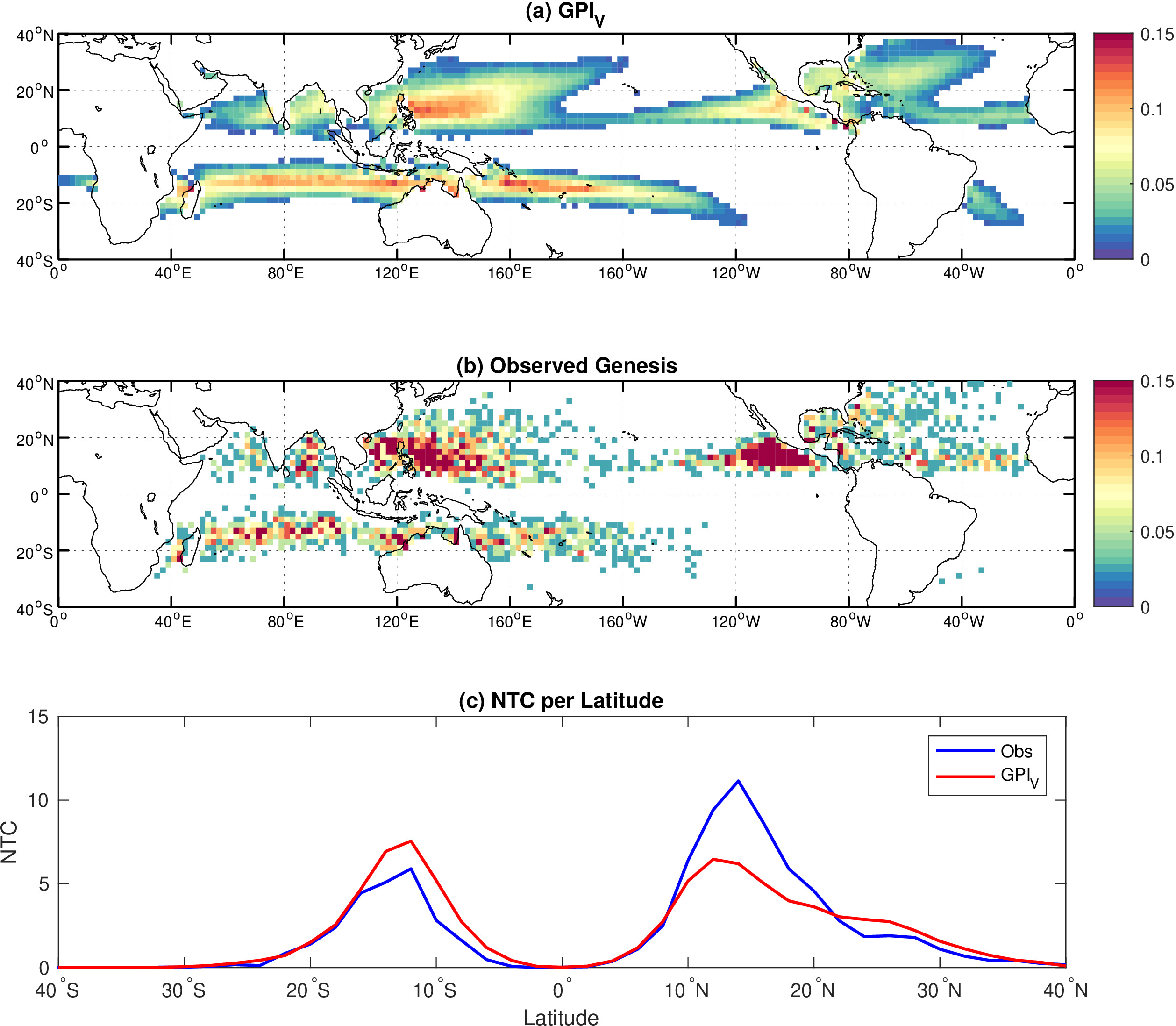}\\
 \caption{Maps of annual-mean (a) $\text{GPI}_V$ (function of $vPI$); (b) observed genesis density [1/yr] 1981-2020, (c) Zonal mean genesis per latitude in observations and integrated $\text{GPI}_V$.}\label{fig:GPIv_genesis}
 \end{center}
\end{figure*}

\begin{figure*}[t]
 \begin{center}
\noindent\includegraphics[width=0.9\textwidth]{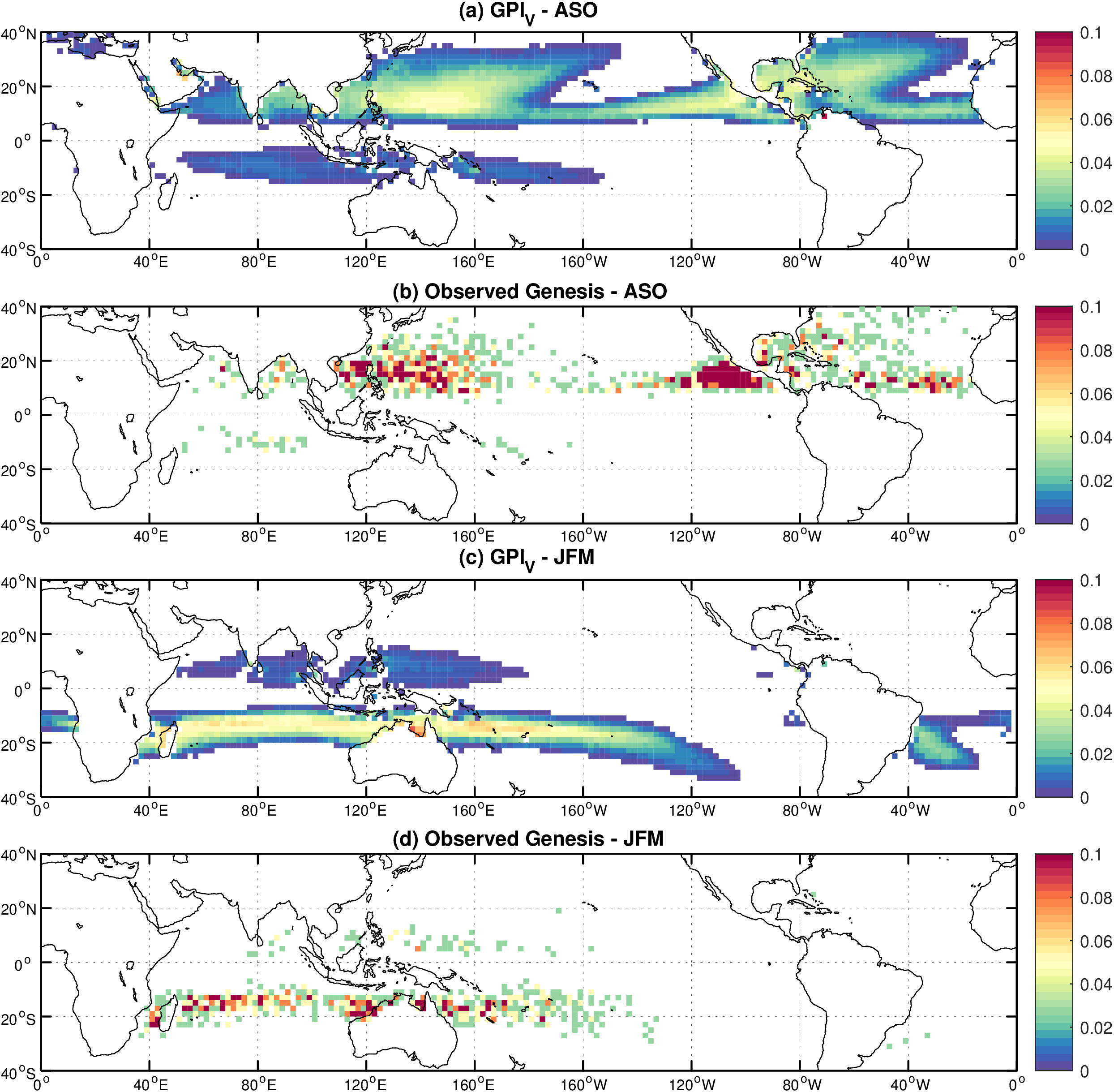}\\
 \caption{Maps of seasonal-mean $\text{GPI}_V$ and observed genesis density [y$^{-1}$] 1981--2020 for (a--b) boreal summer (August to October, ASO); (c--d) austral summer (January to March, JFM). Seasonal values are simply the sum of the GPI over the season, averaged across years.}\label{fig:GPIv_seasonal}
 \end{center}
\end{figure*}

The annual-mean climatology of $GPI_v$ and observed genesis density is shown in Figure \ref{fig:GPIv_genesis}. The spatial distribution of $GPI_v$ (Figure \ref{fig:GPIv_genesis}a) aligns well with that of historical genesis density (Figure \ref{fig:GPIv_genesis}b), capturing the key regions of tropical cyclogenesis globally. $GPI_v$ is largest over the tropical western North Pacific south of 20$^\circ$N between 120$^\circ$E and 150$^\circ$E, as well as over relatively narrow latitudinal bands (10-15$^\circ$N) in the South Indian and Western South Pacific basins west of 170$^\circ$W. In the Eastern North Pacific, which is known to have the highest genesis density in the world, $GPI_v$ is also relatively large though not as large as the aforementioned regions. In the North Atlantic basin, $GPI_v$ is largest over the Gulf of Mexico and western Caribbean, though is overall smaller in magnitude than the other basins. Finally, a small region with modest values of $GPI_v$ is found in the western South Atlantic off the coast of Brazil, consistent with the infrequent and weak tropical cyclones that form in this subregion but not elsewhere in the basin \citep{Pezza_Simmonds_2005,Evans_Braun2012,Gozzo_etal_2014}. Our new genesis index captures the gross meridional structure of genesis (Figure \ref{fig:GPIv_genesis}c), including the greater meridional extent of genesis in the Northern Hemisphere. However, $GPI_v$ predicts that genesis within the tropics, particularly its peak amplitude, ought to be nearly symmetric between the Northern and Southern Hemispheres, in contrast to the nearly doubled peak genesis density in the Northern Hemisphere relative to the Southern Hemisphere. This is a problem common to existing GPIs as well as shown below.


\subsubsection{Annual cycle}

The seasonal-mean climatologies of $GPI_v$ for August-September-October (ASO) and January-February-March (JFM), which correspond to the peak three-month periods of TC activity in each hemisphere, and observed genesis densities for each season are shown in Figure \ref{fig:GPIv_seasonal}. Similar to the annual-mean, $GPI_v$ captures the spatial pattern of TC activity in the peak season of each hemisphere. It also successfully captures the rare occurrence of genesis events in the opposite hemisphere, with low but non-zero values over the tropical South Indian and Southwestern Pacific basins in Boreal summer and over the tropical North Indian Ocean and Northwestern Pacific in Austral summer.

\begin{figure*}[t]
 \noindent\includegraphics[width=0.9\textwidth]{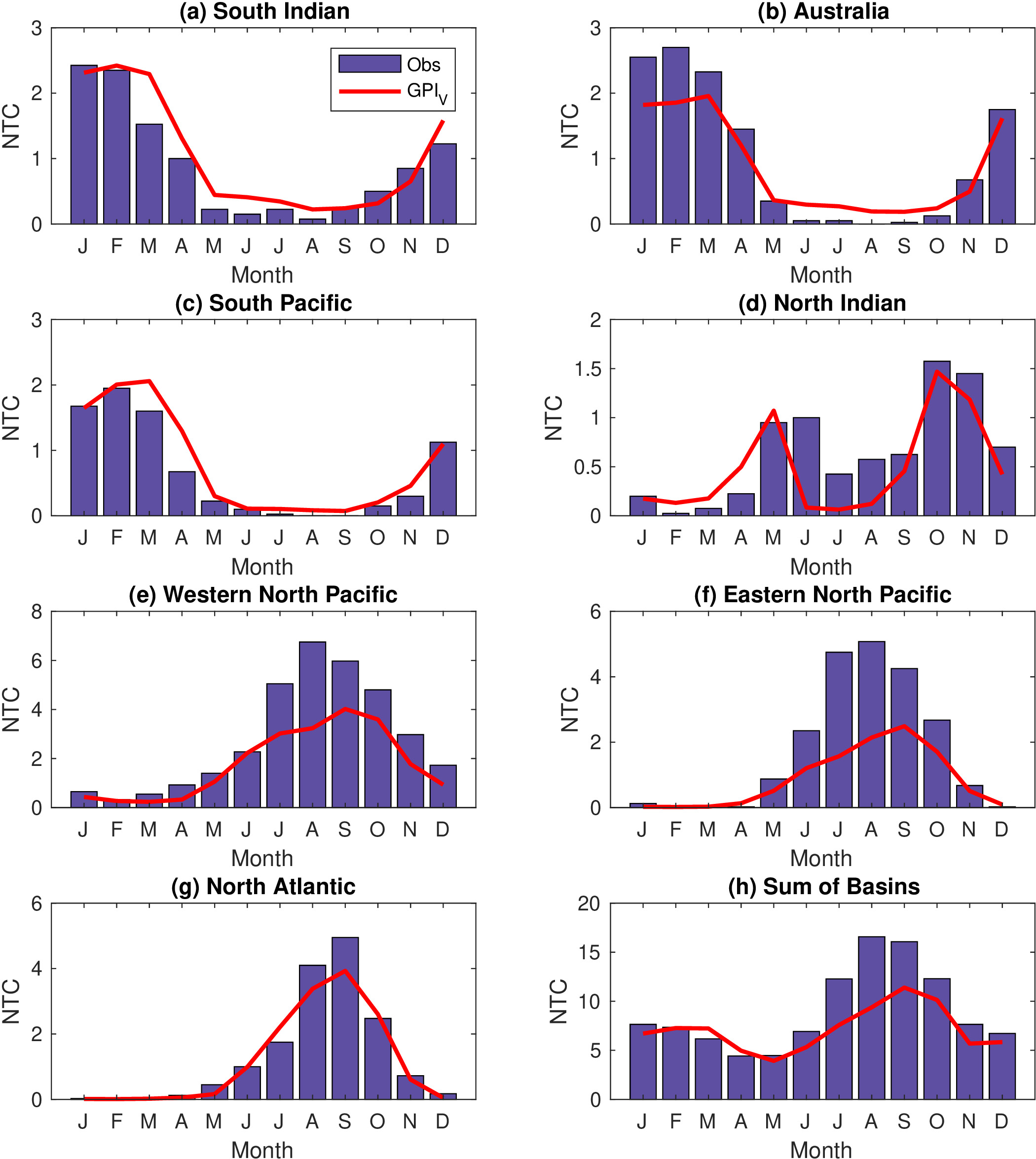}\\
 \caption{Monthly-mean new GPI and storm count by basin (a-g) and the sum across all basins globally (h). For (h), the sum is slightly smaller than the true global count because a small number of genesis events occur outside of basin boundaries.}\label{fig:GPIv_monthlybasins}
\end{figure*}

The annual cycle climatologies of monthly $GPI_v$ and observed genesis count within each basin are shown in Figure \ref{fig:GPIv_monthlybasins}. Overall, $GPI_v$ does reasonably well in reproducing the annual cycle of genesis across all basins and globally. It does very well in the South Indian (Figure \ref{fig:GPIv_monthlybasins}a) and South Pacific (Figure \ref{fig:GPIv_monthlybasins}c), as well as in Australia (Figure \ref{fig:GPIv_monthlybasins}b) and the North Atlantic (Figure \ref{fig:GPIv_monthlybasins}g) though with slight underestimation of genesis counts during the peak months. There is a stronger underestimation of genesis throughout the primary TC season of approximately 20--30\% in the Western North Pacific (Figure \ref{fig:GPIv_monthlybasins}e) and $\sim$ 30--70\% in the Eastern North Pacific (Figure \ref{fig:GPIv_monthlybasins}f). These low-biased basins drive the overall underestimation of genesis count in the Northern Hemisphere noted above and that shows up globally (Figure \ref{fig:GPIv_monthlybasins}g).

\subsubsection{ENSO}

\begin{figure*}[t]
 \noindent\includegraphics[width=0.9\textwidth]{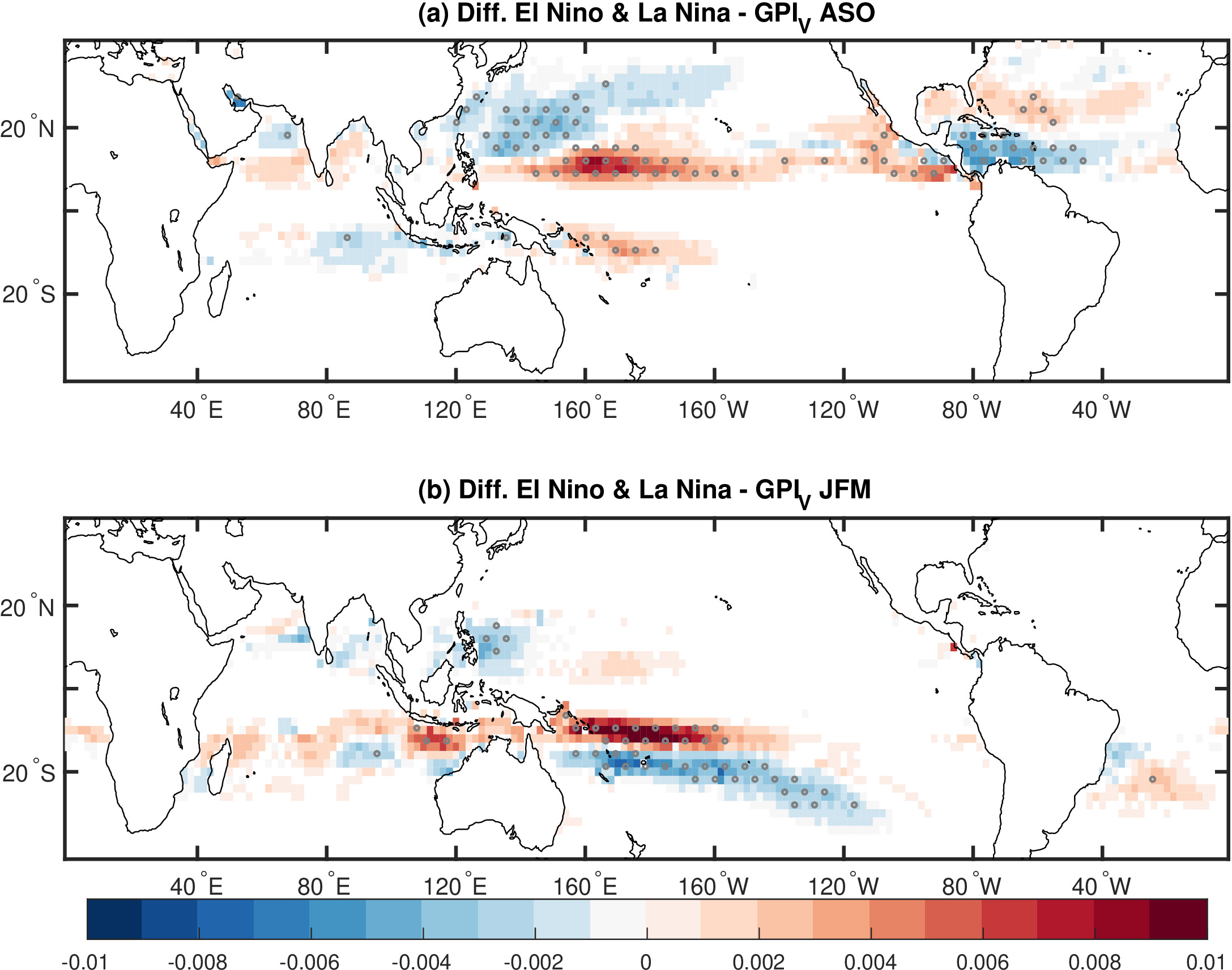}\\
 \caption{Difference of the new GPI anomaly composites between El Ni\~no and La Ni\~na events in the period 1950--2020 in JFM (top) and ASO (bottom). Hatching indicates difference is statistically significant at 95\% confidence level.}
 \label{fig:GPIv_enso}
\end{figure*}

GPIs are known to skillfully represent the variability in TC activity related to the El Ni\~no-Southern Oscillation \citep[ENSO; e.g.,][]{Camargo_Emanuel_Sobel_2007}. Maps of composite differences of $GPI_v$ between El Ni\~no and La Ni\~na events for JFM and ASO are shown in Figure \ref{fig:GPIv_enso} for the period 1950--2020. In  El Ni\~no events, $GPI_v$ is reduced over the tropical Atlantic, particularly through the western half of the Main Development Region where most TCs form, while it is modestly enhanced over the subtropical North Atlantic. Meanwhile, $GPI_v$ is enhanced throughout most of the Eastern North Pacific; this shift in genesis between North Atlantic and East Pacific is a well-documented response to ENSO \citep{Camargo_Emanuel_Sobel_2007}. Additionally, $GPI_v$ is larger across most of the central and western tropical North Pacific, with smaller reductions in the far western tropical and subtropical Pacific.  For the southern Hemisphere, $GPI_v$ is enhanced over the tropical western South Pacific to the east of New Zealand while it is reduced over the tropical eastern South Indian Ocean to the west of southern Indonesia and northern Australia.

\subsubsection{Dependence on common physical variables}

\begin{figure*}[t]
 \begin{center}
 \noindent\includegraphics[width=0.75\textwidth]{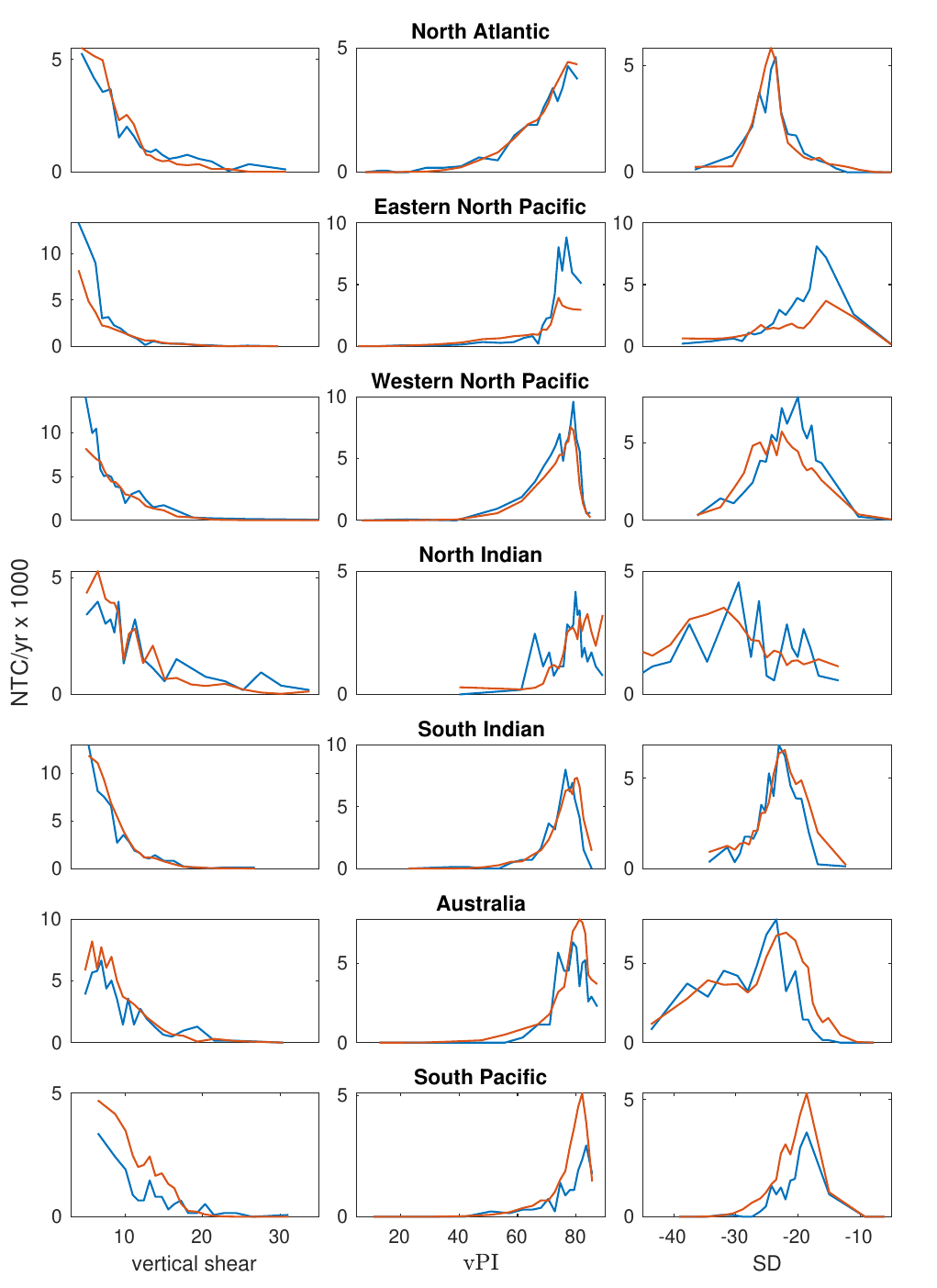}\\
 \end{center}
 \caption{Dependence of observations (blue) and  $GPI_v$ (red) on 850--200 hPa vertical shear, PI, and saturation deficit (SD) by basin.
 }\label{fig:GPIv_paramdep}
\end{figure*}

Finally, we provide a deeper analysis of the performance of $GPI_v$ within each basin by examining how well it captures the observed dependence of TC genesis on related physical parameters in Figure \ref{fig:GPIv_paramdep}: 200-850 hPa vertical wind shear (Figure \ref{fig:GPIv_paramdep}a), ventilated potential intensity (here $vPI$; Figure \ref{fig:GPIv_paramdep}b), and saturation deficit (Figure \ref{fig:GPIv_paramdep}c). This analysis complements the analysis of the spatial and temporal variations in the previous sections by now testing the dependence on physical parameters themselves across their full range of observed values. Note that the dependence on observed genesis count shown in Figure \ref{fig:GPIv_paramdep} (blue curves) can be used to directly infer the ranges of parameter values that are most important for TC activity on Earth: most TCs form with low vertical wind shear (left column), high ventilated potential intensity (middle column), and moderate saturation deficit (right column).

\begin{figure*}[t]
\noindent\includegraphics[width=0.8\textwidth]{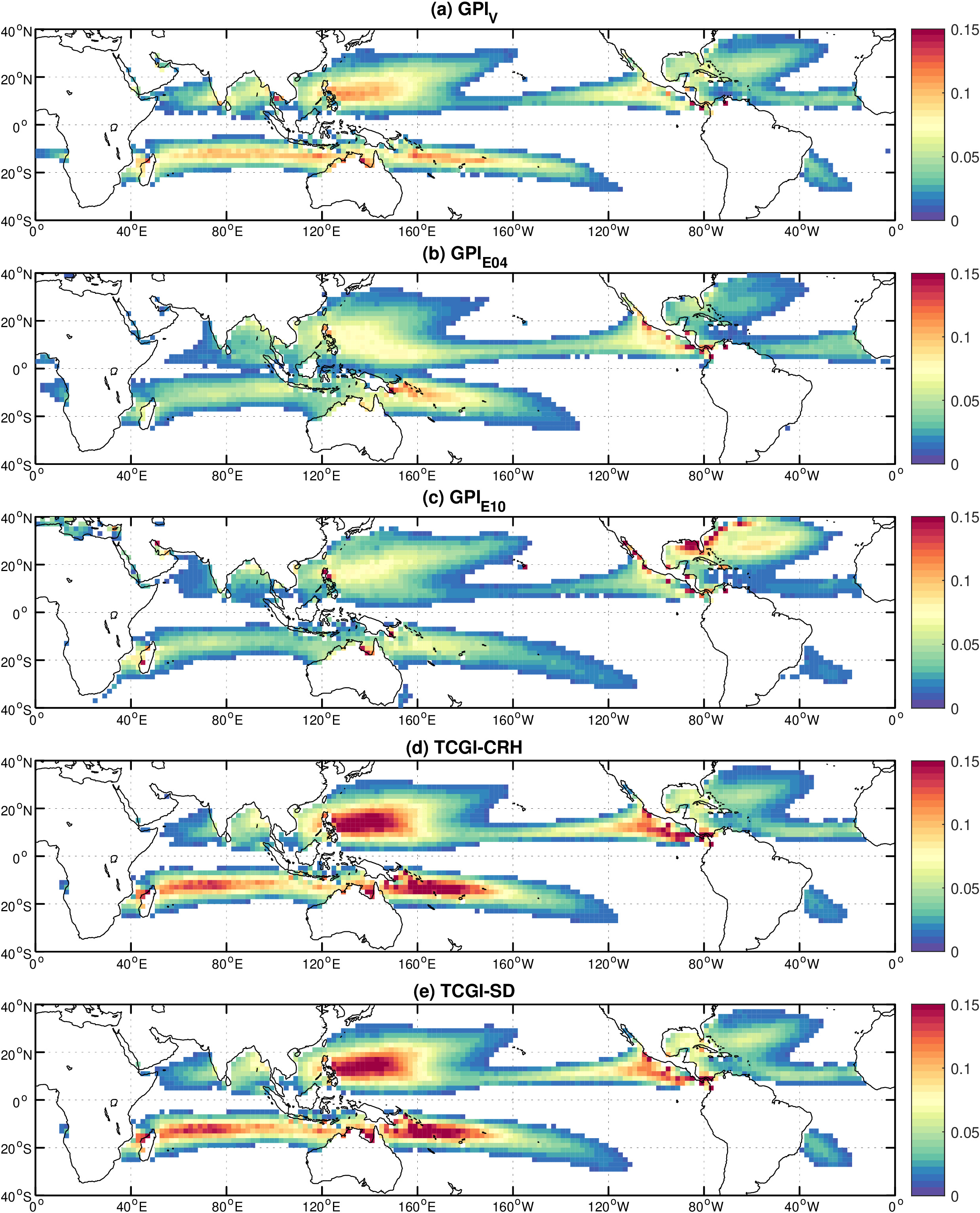}\\
\centering
 \caption{Annual mean climatology for the period 1981-2020 for (a) GPI$_V$, (b) Emanuel's original index (GPI$_{EO4}$)), (c) Emanuel's 2010 index (GPI$_{E10}$), (d) TCGI-CRH and (e) TCGI-SD for the ERA5 reanalysis regrided to a 2 deg. by 2 deg. grid. Emanuel's indices were normalized to match the other indices. Values below $5 \times 10^-3$ are set to zero (white). All indices integrate globally to a value of 84.625 $yr^{-1}$.}
 \label{fig:GPI_comparemaps}
\end{figure*}

In the North Atlantic, South Indian, and Australian basins, the dependence for $GPI_v$ closely matches the dependence for observed genesis count on all three parameters. The dependence of $GPI_v$ is biased low across all three variables in the Eastern North Pacific and Western North Pacific basins, and it is biased high across all three variables in the South Pacific basins. Meanwhile, in the North Indian basin the biases are mixed, though with a high bias at low vertical shear and high $vPI$. Hence, overall our $GPI_v$ are capturing these physical dependencies well, but the weaker than observed dependence in the North Pacific indicates an important gap not captured by our genesis potential index.

\subsection{Comparison with existing GPIs}

\begin{figure*}[t]
\noindent\includegraphics[width=\textwidth,trim={0 3cm 0 0},clip]{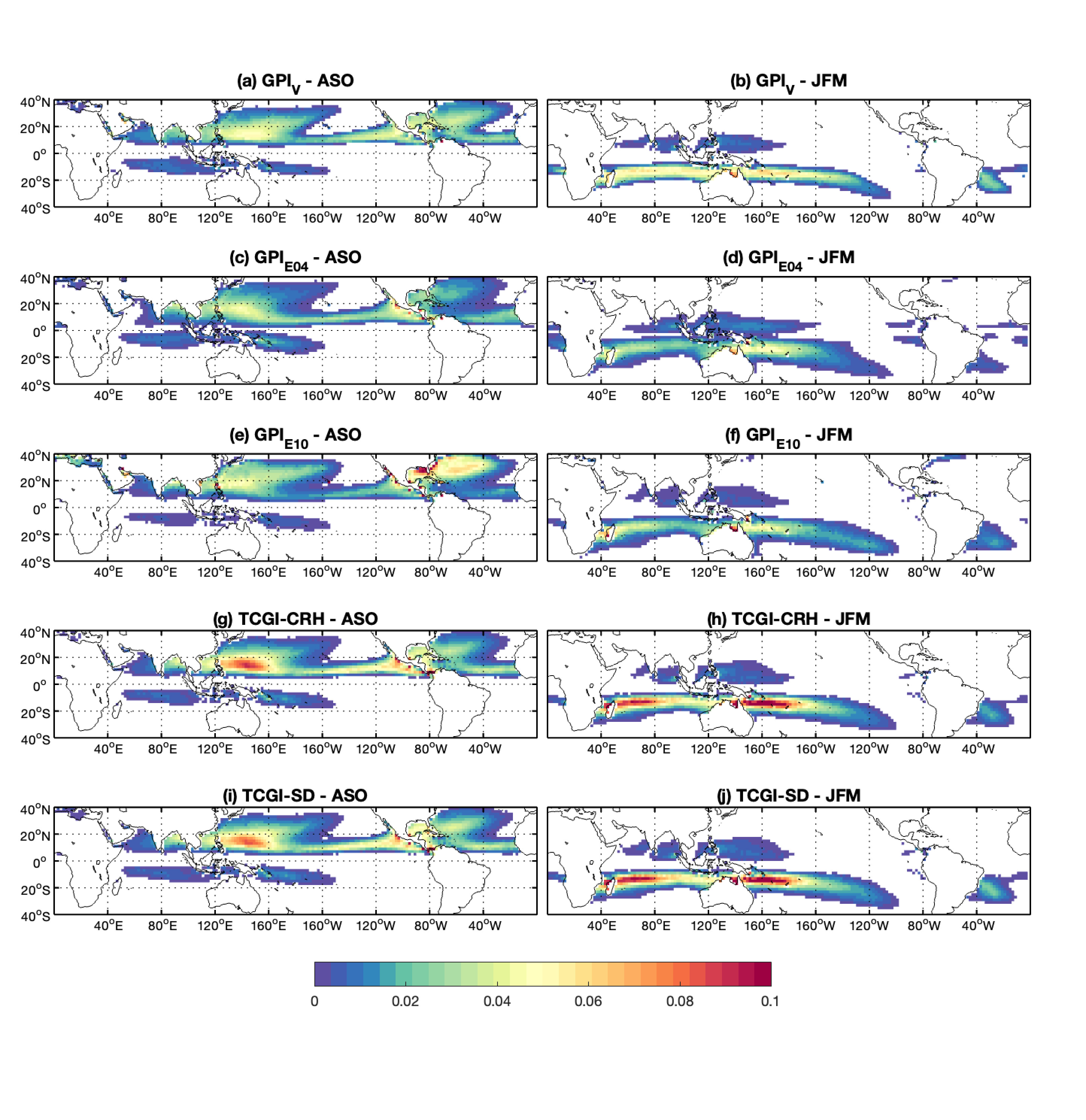}\\
 \caption{Seasonal mean climatology in ASO and JFM for GPI$_V$ (a) and (b), GPI$_{E04}$ (c) and (d), GPI$_{E10}$ (e) and (f), TCGI-CRH (g) and (h) and TCGI-SD (i) and (j) for the ERA5 reanalysis regrided to a 2 deg. by 2 deg grid. Emanuel's indices were normalized to match the other indices.}
 \label{fig:GPI_compareseasonal}
\end{figure*}

The annual-mean climatology of $GPI_v$ and the other four GPI formulations are shown in Figure \ref{fig:GPI_comparemaps}, with seasonal climatologies shown in Figure \ref{fig:GPI_compareseasonal}. The overall spatial distribution of $GPI_v$ is quite similar to the other formulations. The differences between $GPI_v$ and existing GPIs are not obviously larger than among the existing GPI formulations. Note that the combination of shear and relative humidity or saturation deficit in existing GPIs serve to greatly reduce GPI over the marginal parts of the ocean basins in a similar manner as our $vPI$ does in a single parameter. Ultimately, we are not arguing that our spatial distribution of $GPI_v$ is superior to the others, but rather that it performs comparably well to the others in a more parsimonious and physically-grounded manner.

\begin{figure*}[t]
\centering
\noindent\includegraphics[width=0.8\textwidth]{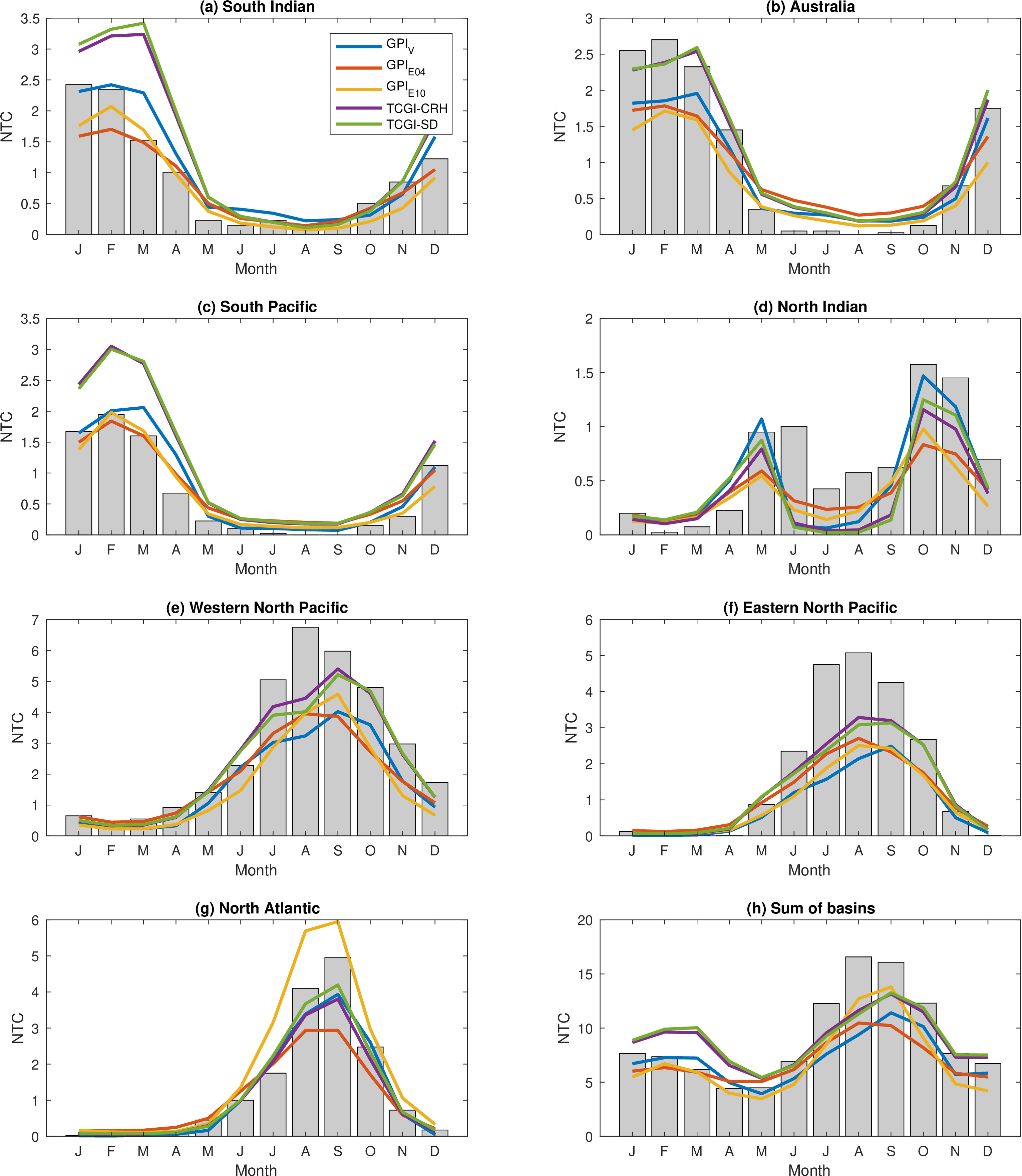}\\
 \caption{Annual cycle of all indices (lines: GPI$_V$ in blue, GPI$_{E04}$ in orange, GPI$_{E10}$ in yellow, TCGI-CRH in purple, and TCGI-SD in green)  and observations (grey bars) in different basins and the sum of all basins.}
 \label{fig:GPI_comparemonthlybasins}
\end{figure*}

The annual cycle climatologies comparing $GPI_v$ to the other GPI formulations within each basin and globally are shown in Figure \ref{fig:GPI_comparemonthlybasins}. Clearly all GPI formulations successfully capture the main regions of TC activity globally, but the relative magnitudes between basins differ. Our $GPI_v$ falls somewhere in between the Emanuel GPIs, which tend to show highest values in the East Pacific and smaller values in the West Pacific (and mixed outcomes in the North Atlantic), and the TCGIs, which show the highest values in the West Pacific, moderate values in the East Pacific, and lower values in the North Atlantic. Across GPIs, the two based on either the normalized entropy deficit (E10) or the saturation deficit (TCGI-SD) tend to show higher magnitudes over the North Atlantic relative to the East Pacific, an outcome that is also evident in our $GPI_v$ in line with their shared physical foundation. Ultimately, though, none are conclusively better or worse than the others. However, our $GPI_v$ is more parsimonious in form than the others and rooted more strongly in tropical cyclone physics, and hence may provide clear guidance on the representing the free-tropospheric moisture variable and its impact on the spatial distribution of GPIs in their existing formulations.

\subsection{Future projections in CMIP6}

\begin{figure*}[t]
\centering
\noindent\includegraphics[width=0.9\textwidth]{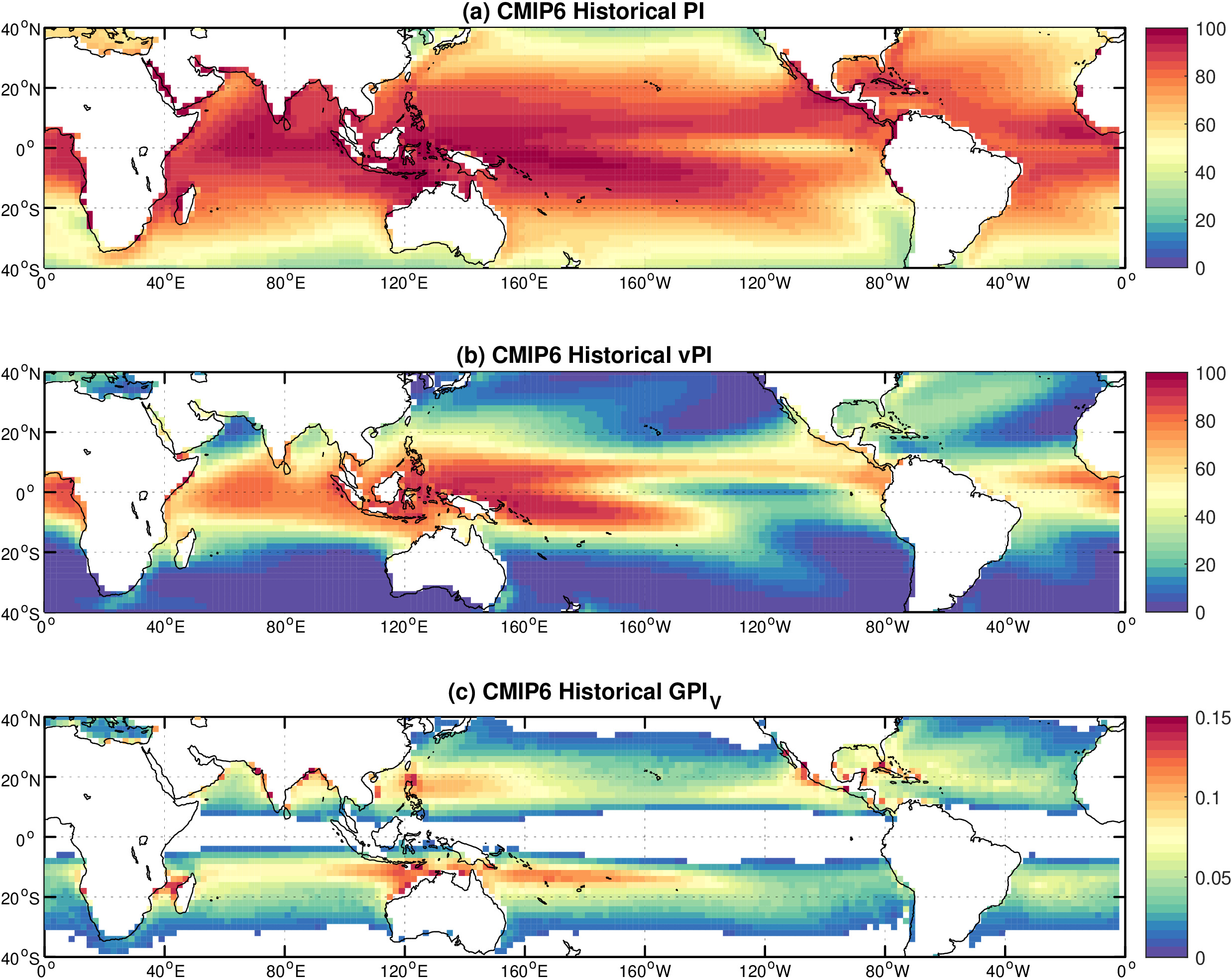}\\
 \caption{Historical (1981 - 2010) multi-model-mean annual climatology for 45 CMIP6 models for  (a) potential intensity, (b) ventilation reduced potential intensity and  (c) GPI$_V$ (bottom).}
 \label{fig:GPIv_components_cmip}
\end{figure*}

Finally, we examine future projected changes in $GPI_v$ from the CMIP6 multi-model ensemble. We begin first with the multi-model mean historical climatology of $PI$, $vPI$, and $GPI_v$ in Figure \ref{fig:GPIv_components_cmip}. The spatial distribution closely aligns with that found in reanalysis in Figure \ref{fig:GPIv_genesis} above. Accounting for ventilation in the potential intensity again yields a much more refined geographic distribution of environments favorable for TCs. The distribution of $GPI_v$ aligns with the key regions of TC activity, with a broader region of very small values extending into higher latitudes relative to ERA5 owing to model biases and the smoothing that arises from averaging over many ensembles and models in those regions.


\begin{figure*}[t]
\noindent\includegraphics[clip=true, trim={2.5cm 1.3cm 3cm 0.08cm},width=\textwidth]{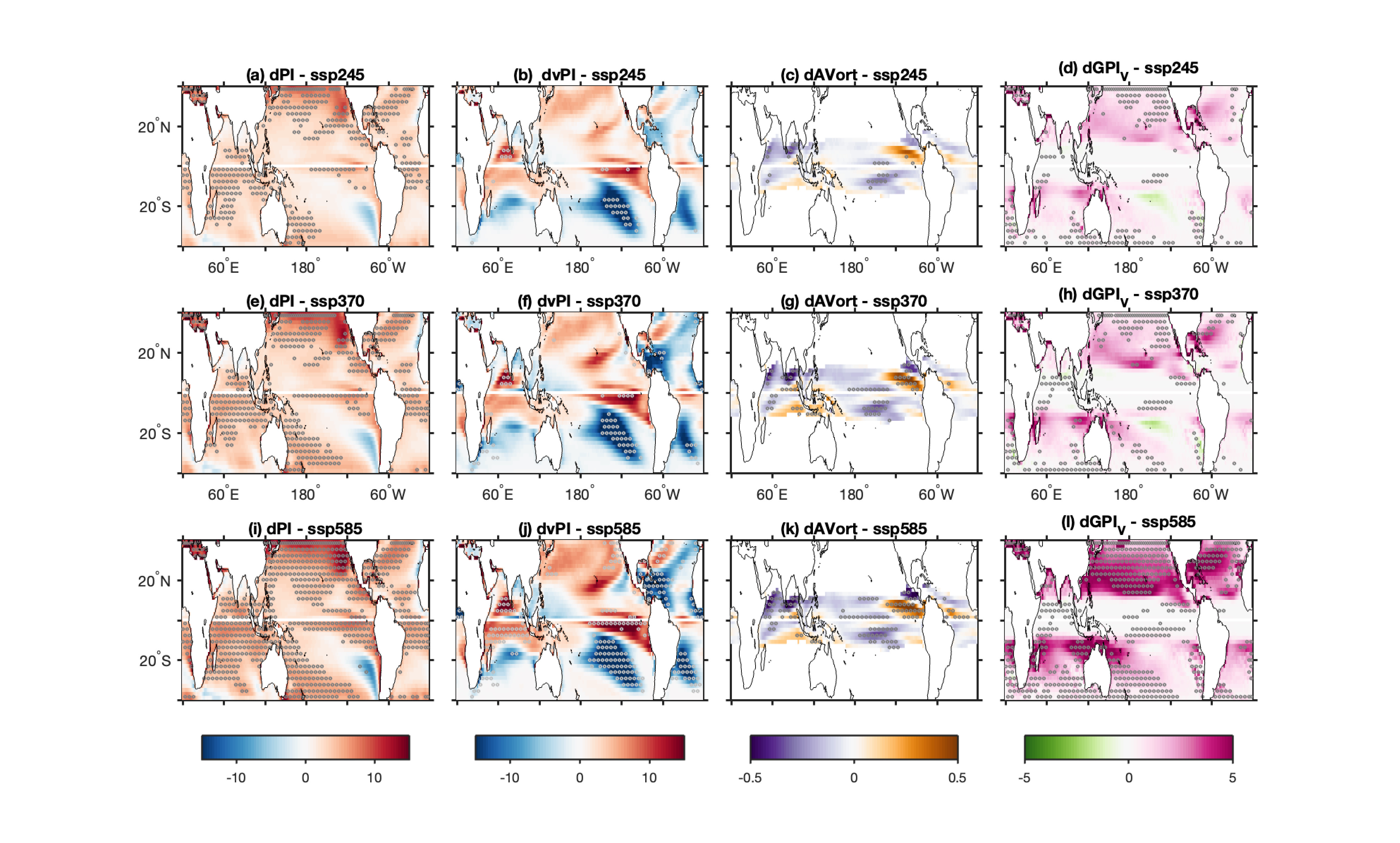}\\
 \caption{Differences between the CMIP6 multi-model mean (Northern Hemisphere for August to October, southern hemisphere for January to March) at the end of the 21C century (2071-2100) and end of 20C (1971-2000) for $PI$ (a,e,i), $vPI$ (b,f,j), $\eta_c \times 10^5$ (c,g,k), and GPI$_V \times 10^3$ (d,h,l) for 3 different future scenarios ssp245 (top panels, 40 models), ssp370 (middle panels, 36 models), ssp585 (45 models). Statistically significant differences among the multi-model ensemble based on the Kolmogorov-Smirnov test at the 95\% level are marked with a circle.}
 \label{fig:GPIv_diff_cmip}
\end{figure*}

Figure \ref{fig:GPIv_diff_cmip} displays the differences in the multi-model mean climatology of $PI$, $vPI$, $\eta_c$, and $GPI_v$ for ssp245, ssp370, and ssp585 between the end of the 21C (2071--2100) and the end of the 20C (1971--2000). Basin-integrated changes in $GPI_v$ are shown in Table \ref{tab:percGPIv}. Globally there is a consistent broad increase in $GPI_v$ across all basins (4th column), with the magnitude increasing with increasing warming across scenarios while maintaining a very similar spatial pattern (rows). The eastern and central North Pacific, subtropical North Atlantic, and South Indian ocean basins are regions of largest increase, with the subtropical western North Pacific showing slightly weaker increases. Changes in $GPI_v$ are driven almost entirely by changes in $vPI$ (2nd column), particularly at higher latitudes, with increases of 5-15 $ms^{-1}$ in the aforementioned regions. Because the absolute vorticity $\eta_c$ is capped at $3.7\times 10^{-5} \: s^{-1}$, which corresponds to an equivalent Coriolis latitude of approximately 15 degrees, and this term is generally dominated by $f$, changes in its impact are effectively limited to within the deep tropics (3rd column). Changes in $\eta_c$ within the deep tropics have a very small impact on $GPI_v$ with the exception of a slight increase in the near-equatorial eastern North Pacific basin and slight decrease in the near-equatorial North Indian basin, the latter offset by an increase in $vPI$. Changes in $vPI$ broadly follow the changes in $PI$ (left column), particularly the increase in both poleward of 20$^\circ$N. However, within the tropics there are regions where the two diverge, partiuclarly the western Atlantic and Pacific basins and central South Indian basins where $vPI$ decreases despite an increase in $PI$. Note that there is a strong decrease in mean $vPI$ in the central South Pacific and tropical western Atlantic, yet $GPI_v$ decreases only slightly in the former while remaining relatively constant in the latter, with $GPI_v$ still consistently increasing in the Atlantic basin as a whole (Table \ref{tab:percGPIv}). This outcome highlights that the $GPI_v$ is a non-linear combination of $vPI$ and $\eta_c$ and hence mean changes in $GPI_v$ may not always neatly follow mean changes in the individual components.

Comparing results across scenarios and time periods, the response is quantitatively very similar between ssp245 and ssp370, with only a slightly higher increase across basins in the latter by the end of the century owing to the stronger global warming by that period. The patterns exhibit similar localized regions of statistically-significant increases in the Northern Hemisphere subtropics (Figure \ref{fig:GPIv_components_cmip}d vs.\ h). The basin-wide response magnitude within each basin scales closely with global-mean warming (Table \ref{tab:percGPIv}), with the largest responses by the end of the 21st century when warming is strongest.  The signal of increasing $GPI_v$ at higher latitudes in the subtropical North Atlantic and North Pacific emerges by end of century across all scenarios, but it becomes much more pronounced in both hemispheres only in ssp585, corresponding to the greatest magnitude of warming. The end-of-century ssp585 scenario exhibits a substantially larger jump in $GPI_v$ relative to the global-mean warming, with percentage increases in $GPI_v$ nearly double that of ssp370 despite a global-mean warming only 20\% larger ($4.9 \: ^\circ$C vs. $4.0 \: ^\circ$C). Recall that ssp245 is the closest analog to our current emissions trajectory, with ssp370 representing a reasonable high-end scenario that is considered more plausible than ssp585 \citep{Pielke_Burgess_Ritchie_2022}.


\begin{table*}[t]
\caption{Percentage changes of mean $GPI_{v}$ in the tropics in different basins, southern hemisphere basins in JFM, northern hemisphere basins in ASO for 3 periods: 2021-2040 (P1), 2041-2060 (P2) and 2071-2100 (P3). Also shown are the estimated multi-model mean global mean surface temperature increases at the mid-point of each period (2030, 2050, and 2085)}
\begin{center}
\begin{tabular}{|c|ccc|ccc|ccc|}
\hline\hline
 & \multicolumn{3}{|c|}{ssp245} & \multicolumn{3}{|c|}{ssp370} & \multicolumn{3}{|c|}{ssp585} \\ \hline
Basin & P1 & P2 & P3 & P1 & P2 & P3 & P1 & P2 & P3 \\ \hline
SI & 7.6\% & 10.8\% & 16.2\% & 
7.2\% & 10.7\% & 20.9\% & 
11.5\% & 18.0\% & 45.9\% \\
AUS & 6.3\% & 8.9\% & 13.9\% & 
5.3\% & 7.5\% & 17.4\% & 
8.7\% & 14.2\% & 37.9\% \\
SP & 1.1\% & 0.0 & 2.0\% &
2.5\% & 2.0\% & 6.6\% &
4.9\% & 7.9\% & 22.1\% \\
WNP & 5.8\% & 9.8\% & 15.5\% &
4.2\% & 8.0\% & 17.2\% & 
8.9\% & 16.5\% & 41.9\% \\
ENP & 12.3\% & 16.9\% & 23.1\% &
12.9\% & 16.9\% & 28.1\% &
20.3\% & 29.4\% & 60.4\% \\
ATL & 14.1\% & 16.5\% & 22.6\% &
12.4\% & 15.9\% & 24.7\% &
23.1\% & 31.1\% & 61.6\% \\ \hline
TS & 1.6$^\circ$C & 2.2$^\circ$C & 3.0$^\circ$C &
1.6$^\circ$C & 2.3$^\circ$C & 4.0$^\circ$C & 
1.7$^\circ$C & 2.6$^\circ$C & 4.9$^\circ$C \\
\hline
\end{tabular}
\end{center}
\label{tab:percGPIv}
\end{table*}

Overall, our results indicate that the global environment becomes very gradually more favorable for tropical cyclones in general (as measured by $vPI$) and for genesis in particular (as measured by $GPI_v$), though these changes are relatively modest for current plausible scenarios through the end of 2100. The first signal to emerge is an increase in higher-latitude genesis as warming magnitude increases, whether due to higher emissions or more time at lower emissions, which is consistent with recent poleward trends in TC activity, in particular in the western North Pacific \citep{Kossin_Emanuel_Vecchi_2014,Daloz_Camargo2018,Studholme_etal_2022}, as well as future projections \citep{Kossin_etal_2016}, however their could be some compensating effects due to higher latitude environmental characteristics \citep{Lin_etal_2023}. Furthermore, recent trends in TC frequency have been decreasing globally \citep{Klotzbach_etal_2020}. Long term estimates of trends in TC frequency have been a source significant debate in the scientific community, given the uncertainty of the data pre-satellite era and the multiple methods used to make such estimates leading to contrasting results \citep{Vecchi_etal_2021,Emanuel2021,Chand_etal_2022,Emanuel2023,Chand_etal_2023}.

Our new GPI using the ventilated potential intensity predicts a modest increase in TC genesis with warming, consistent with the predictions from the GPI of \cite{Emanuel_2021} based on the normalized entropy deficit, which is the closest GPI cousin to ours. It is also consistent with GPIs that use either mid-level relative humidity or column relative humidity but not column saturation deficit, which might not be expected on intuitive grounds. Hence this speaks to the importance of using physical parameters that fully exploit our underlying theory, as the precise nature of its influence appears to determine its influence in a warming climate.

At the same time, our new GPI is again not a closed theory, and hence the above also further emphasizes the intrinsic uncertainty associated with GPIs of any form as proxies for tropical cyclogenesis. Moreover, it is important to acknowledge ongoing uncertainties in both historical tropical cyclone data and reanalysis data used to train our GPI, and perhaps more importantly in climate models whose future projections for a more El Ni\~{n}o-like pattern are currently sharply at odds with the recent La Ni\~{n}a-like trend \citep{Seager_etal_2019,Sobel_etal_2023}. Given the relatively modest increases in our GPI with warming, these changes may certainly be viewed as well within the combined error bars of our estimation of the current climate and projection of future climate. As such, our work may help settle the debate regarding the treatment of moisture within a GPI, but it cannot conclusively settle the broader debate about whether genesis itself will change with warming.


\section{Conclusions}

This work has developed a novel genesis potential index that depends on a single parameter given by the product of the ventilated potential intensity and the clipped absolute vorticity. The ventilated potential intensity represents the thermodynamic favorability of an environment for supporting a tropical cyclone by explicitly incorporates the detrimental effect of mid-level low-entropy environmental air on the potential intensity and is taken directly from existing ventilation theory. This quantity was developed and successfully applied in recent work to understand tropical cyclogenesis on tidally-locked exoplanets, but it had yet to be applied to Earth. The new GPI is more parsimonious than existing indices while performing comparably well in reproducing the climatological distribution of tropical cyclone activity and its covariability with ENSO. When applied to CMIP6 projections, the new GPI predicts that environments globally will become gradually more favorable for TC activity (as measured by the ventilated potential intensity) and genesis (GPI) with warming. However, significant changes emerge only under relatively strong warming and principally at higher latitudes outside of the tropics. This qualitative outcome is consistent with the projected increases found by \cite{Emanuel_2021} using a GPI that takes the normalized entropy deficit as its environmental moisture predictor, whose formulation is motivated by the same underlying theory that we have used explicitly in creating our new GPI presented here. Hence, our results emphasize the importance of employing parameters whose form are fully rooted in theory. Our findings should help resolve the debate over the treatment of the moisture term in a GPI and the implication of this choice for how TC genesis may change with warming.

It is important to emphasize that genesis potential indices, including our own, remain semi-empirical proxies for TC genesis that carry intrinsic uncertainty in extrapolating to future climate states owing to the nature of their formulation in the absence of a complete theory for genesis. However, the core parameter in our new GPI is given by the simple product of the ventilated potential intensity, which is the fundamental thermodynamic ingredient for the energetics of the TC, and the absolute vorticity, which is the fundamental dynamic ingredient for the rotation of the TC. This joint parameter has units of acceleration, which could plausibly be interpreted as analogous to an intensification rate. Hence, this parameter carries deep intuitive physical appeal as a crude representation of the growth rate of a weak initial disturbance. Why the parameter should be taken to the fifth power has no intuitive explanation, though, and instead is an indication of how this remains well short of a theory for genesis. Instead, our new GPI may perhaps be viewed as the most refined possible representation of two necessary (but not sufficient) ingredients for tropical cyclogenesis: energy and rotation. Clearly, localized and sustained upward motion in the form of a ``seed'' is also a necessary condition \citep{Merlis_Zhao_Held_2013,Sugietal2020,Hsieh_eatl_2020,Zhang_etal_2021b}, whose absence in GPIs likely explains why they struggle to predict changes in TC genesis on sub-climatological timescales \citep{Cavicchia_etal_2023}. Our GPI provides no further insight on this matter relative to previous iterations. However, the physical simplicity and parsimony of our version may provide a useful step towards a broader theory of genesis that is skillful in predictions across all timescales, including future climate change.




%

%

\clearpage
\acknowledgments

DRC was supported by NSF AGS grant 1945113. SJC was supported by NSF grants AGS 20-43142, 22-17618, 22-44918 and NOAA grants NA21OAR4310345 and NA22OAR4310610.  


%
%
\datastatement
IBTrACS data are publicly available from the National Center for Environmental Information website at \url{https://www.ncei.noaa.gov/products/international-best-track-archive}. ERA5 reanalysis data is publicly available from the European Center for Medium Range Weather Forecasting Copernicus Climate Data Store at \url{https://cds.climate.copernicus.eu/}. Coupled Model Intercomparison Project version 6 model simulation data are publicly available at \url{https://aims2.llnl.gov/search/cmip6/}. 


%






%



\bibliographystyle{ametsocV6}
\bibliography{refs_CHAVAS,Chavas_CMIP6}

\newcommand{\noopsort}[1]{} \newcommand{\printfirst}[2]{#1} \newcommand{\singleletter}[1]{#1} \newcommand{\switchargs}[2]{#2#1}
\begin{thebibliography}{109}
\providecommand{\natexlab}[1]{#1}
\providecommand{\url}[1]{\texttt{#1}}
\renewcommand{\UrlFont}{\rmfamily}
\providecommand{\urlprefix}{URL }
\expandafter\ifx\csname urlstyle\endcsname\relax
  \providecommand{\doi}[1]{https://doi.org/\discretionary{}{}{}#1}\else
  \providecommand{\doi}{https://doi.org/\discretionary{}{}{}\begingroup \urlstyle{rm}\Url}\fi
\providecommand{\eprint}[2][]{\url{#2}}

\bibitem[{Andrews et~al.(2019)}]{andrewsForcingsFeedbacksClimate2019}
Andrews, T., and Coauthors, 2019: Forcings, {{Feedbacks}}, and {{Climate Sensitivity}} in {{HadGEM3-GC3}}.1 and {{UKESM1}}. \textit{J. Adv. Model. Earth Syst.}, \textbf{11~(12)}, 4377--4394, \doi{10.1029/2019MS001866}.

\bibitem[{Bao et~al.(2020)Bao, Song,, and Qiao}]{baoFIOESMVersionModel2020}
Bao, Y., Z.~Song, and F.~Qiao, 2020: {{FIO-ESM Version}} 2.0: {{Model Description}} and {{Evaluation}}. \textit{J. Geophys. Res.: Oceans}, \textbf{125~(6)}, e2019JC016\,036, \doi{10.1029/2019JC016036}.

\bibitem[{Bi et~al.(2020)}]{biConfigurationSpinupACCESSCM22020}
Bi, D., and Coauthors, 2020: Configuration and spin-up of {{ACCESS-CM2}}, the new generation {{Australian Community Climate}} and {{Earth System Simulator Coupled Model}}. \textit{J. South. Hemisphere Earth Syst. Sci.}, \textbf{70}, 225--251, \doi{10.1071/ES19040}.

\bibitem[{Bister and Emanuel(2002)Bister, and Emanuel}]{Bister_Emanuel_2002}
Bister, M., and K.~Emanuel, 2002: Low frequency variability of tropical cyclone potential intensity. part 1: {I}nterannual to interdecadal variability. \textit{Journal of Geophysical Research}, \textbf{107~(D24)}, 4801.

\bibitem[{Boucher et~al.(2020)}]{boucherPresentationEvaluationIPSLCM6ALR2020}
Boucher, O., and Coauthors, 2020: Presentation and {{Evaluation}} of the {{IPSL-CM6A-LR Climate Model}}. \textit{J. Adv. Model. Earth Syst.}, \textbf{12~(7)}, e2019MS002\,010, \doi{10.1029/2019MS002010}.

\bibitem[{Bretherton et~al.(2004)Bretherton, Peters,, and Back}]{Bretherton_etal_2004}
Bretherton, C.~S., M.~E. Peters, and L.~E. Back, 2004: Relationships between {{Water Vapor Path}} and {{Precipitation}} over the {{Tropical Oceans}}. \textit{Journal of Climate}, \textbf{17~(7)}, 1517--1528, \doi{10.1175/1520-0442(2004)017<1517:RBWVPA>2.0.CO;2}.

\bibitem[{Bruy\`ere et~al.(2012)Bruy\`ere, Holland,, and Towler}]{Bruyere_Holland_Towler2012}
Bruy\`ere, C.~L., G.~J. Holland, and E.~Towler, 2012: Investigating the use of a genesis potential index for tropical cyclones in the north atlantic basin. \textit{Journal of Climate}, \textbf{25}, 8611 -- 8626, \doi{10.1175/JCLI-D-11-00619.1}.

\bibitem[{Camargo et~al.(2007{\natexlab{a}})Camargo, Sobel, Barnston,, and Emanuel}]{Camargo_etal_2007}
Camargo, S., A.~H. Sobel, A.~G. Barnston, and K.~A. Emanuel, 2007{\natexlab{a}}: Tropical cyclone genesis potential index in climate models. \textit{Tellus A: Dynamic Meteorology and Oceanography}, \textbf{59~(4)}, 428--443.

\bibitem[{Camargo(2013)}]{Camargo_2013}
Camargo, S.~J., 2013: Global and regional aspects of tropical cyclone activity in the {CMIP5} models. \textit{Journal of Climate}, \textbf{26~(24)}, 9880--9902.

\bibitem[{Camargo et~al.(2007{\natexlab{b}})Camargo, Emanuel,, and Sobel}]{Camargo_Emanuel_Sobel_2007}
Camargo, S.~J., K.~A. Emanuel, and A.~H. Sobel, 2007{\natexlab{b}}: Use of a genesis potential index to diagnose {ENSO} effects on tropical cyclone genesis. \textit{Journal of Climate}, \textbf{20~(19)}, 4819--4834.

\bibitem[{Camargo et~al.(2014)Camargo, Tippett, Sobel, Vecchi,, and Zhao}]{Camargo_etal_2014}
Camargo, S.~J., M.~K. Tippett, A.~H. Sobel, G.~A. Vecchi, and M.~Zhao, 2014: Testing the performance of tropical cyclone genesis indices in future climates using the {HIRAM} model. \textit{Journal of Climate}, \textbf{27~(24)}, 9171--9196.

\bibitem[{Camargo et~al.(2009)Camargo, Wheeler,, and Sobel}]{Camargo_Wheeler_Sobel_2009}
Camargo, S.~J., M.~C. Wheeler, and A.~H. Sobel, 2009: Diagnosis of the mjo modulation of tropical cyclogenesis using an empirical index. \textit{Journal of the Atmospheric Sciences}, \textbf{66~(10)}, 3061--3074.

\bibitem[{Camargo and Wing(2015)Camargo, and Wing}]{Camargo_Wing_2016}
Camargo, S.~J., and A.~A. Wing, 2015: Tropical cyclones in climate models. \textit{Wiley Interdisciplinary Reviews: Climate Change}.

\bibitem[{Camargo et~al.(2023)}]{Camargo_etal_2023}
Camargo, S.~J., and Coauthors, 2023: An update on the influence of natural climate variability and anthropogenic climate change on tropical cyclones. \textit{Tropical Cyclone Research and Review}, \textbf{12}, 216--239, \doi{10.1016/j.tcrr.2023.10.001}.

\bibitem[{Cao et~al.(2018)}]{caoNUISTEarthSystem2018}
Cao, J., and Coauthors, 2018: The {{NUIST Earth System Model}} ({{NESM}}) version 3: Description and preliminary evaluation. \textit{Geosci. Model. Dev.}, \textbf{11~(7)}, 2975--2993, \doi{10.5194/gmd-11-2975-2018}.

\bibitem[{Cavicchia et~al.(2023)Cavicchia, Scoccimarro, Ascenso, Castelletti, Giuliani,, and Gualdi}]{Cavicchia_etal_2023}
Cavicchia, L., E.~Scoccimarro, G.~Ascenso, A.~Castelletti, M.~Giuliani, and S.~Gualdi, 2023: Tropical cyclone genesis potential indices in a new high-resolution climate models ensemble: Limitations and way forward. \textit{Geophysical Research Letters}, \textbf{50~(11)}, e2023GL103\,001.

\bibitem[{Chand et~al.(2022)}]{Chand_etal_2022}
Chand, S.~S., and Coauthors, 2022: Declining tropical cyclone frequency under global warming. \textit{Nature Climate Change}, \textbf{12~(7)}, 655--661, \doi{10.1038/s41558-022-01388-4}.

\bibitem[{Chand et~al.(2023)}]{Chand_etal_2023}
Chand, S.~S., and Coauthors, 2023: Reply to: {Limitations} of reanalyses for detecting tropical cyclone trends. \textit{Nature Climate Change}, 1--2, \doi{10.1038/s41558-023-01880-5}.

\bibitem[{Chavas(2017)}]{Chavas_2017}
Chavas, D.~R., 2017: A simple derivation of tropical cyclone ventilation theory and its application to capped surface entropy fluxes. \textit{Journal of the Atmospheric Sciences}, \textbf{74~(9)}, 2989--2996.

\bibitem[{Chavas and Reed(2019)Chavas, and Reed}]{Chavas_Reed_2019}
Chavas, D.~R., and K.~A. Reed, 2019: Dynamical aquaplanet experiments with uniform thermal forcing: system dynamics and implications for tropical cyclone genesis and size. \textit{Journal of the Atmospheric Sciences}, \textbf{~(2019)}.

\bibitem[{Daloz and Camargo(2018)Daloz, and Camargo}]{Daloz_Camargo2018}
Daloz, A.~S., and S.~J. Camargo, 2018: Is the poleward migration of tropical cyclone maximum intensity associated with a poleward migration of tropical cyclone genesis? \textit{Climate Dynamics}, \textbf{50~(1-2)}, 705--715, \doi{10.1007/s00382-017-3636-7}.

\bibitem[{Danabasoglu et~al.(2020)}]{danabasogluCommunityEarthSystem2020}
Danabasoglu, G., and Coauthors, 2020: The {{Community Earth System Model Version}} 2 ({{CESM2}}). \textit{Journal of Advances in Modeling Earth Systems}, \textbf{12~(2)}, e2019MS001\,916, \doi{10.1029/2019MS001916}.

\bibitem[{Dirkes et~al.(2023)Dirkes, Wing, Camargo,, and Kim}]{Dirkes_etal_2023}
Dirkes, C.~A., A.~A. Wing, S.~J. Camargo, and D.~Kim, 2023: Process-{Oriented} {Diagnosis} of {Tropical} {Cyclones} in {Reanalyses} {Using} a {Moist} {Static} {Energy} {Variance} {Budget}. \textit{Journal of Climate}, \textbf{36~(16)}, 5293--5317, \doi{10.1175/JCLI-D-22-0384.1}.

\bibitem[{Dunne et~al.(2020)}]{dunneGFDLEarthSystem2020}
Dunne, J.~P., and Coauthors, 2020: The {{GFDL Earth System Model Version}} 4.1 ({{GFDL-ESM}} 4.1): {{Overall Coupled Model Description}} and {{Simulation Characteristics}}. \textit{J. Adv. Model. Earth Syst.}, \textbf{12~(11)}, e2019MS002\,015, \doi{10.1029/2019MS002015}.

\bibitem[{Döscher et~al.(2022)}]{doscherECEarth3EarthSystem2022}
Döscher, R., and Coauthors, 2022: The {{EC-Earth3 Earth}} system model for the {{Coupled Model Intercomparison Project}} 6. \textit{Geosci. Model. Dev.}, \textbf{15~(7)}, 2973--3020, \doi{10.5194/gmd-15-2973-2022}.

\bibitem[{Emanuel(2010)}]{Emanuel_2010}
Emanuel, K., 2010: Tropical cyclone activity downscaled from {NOAA}-{CIRES} reanalysis, 1908-1958. \textit{Journal of Advances in Modeling Earth Systems}, \textbf{2~(1)}, 1--12.

\bibitem[{Emanuel(2021{\natexlab{a}})}]{Emanuel2021}
Emanuel, K., 2021{\natexlab{a}}: Atlantic tropical cyclones downscaled from climate reanalyses show increasing activity over past 150 years. \textit{Nature Communications}, \textbf{12~(1)}, 1--8, \doi{10.1038/s41467-021-27364-8}.

\bibitem[{Emanuel(2021{\natexlab{b}})}]{Emanuel_2021}
Emanuel, K., 2021{\natexlab{b}}: Response of global tropical cyclone activity to increasing co 2: Results from downscaling cmip6 models. \textit{Journal of Climate}, \textbf{34~(1)}, 57--70.

\bibitem[{Emanuel(2023)}]{Emanuel2023}
Emanuel, K., 2023: Limitations of reanalyses for detecting tropical cyclone trends. \textit{Nature Climate Change}, 1--3, \doi{10.1038/s41558-023-01879-y}.

\bibitem[{Emanuel(1986)}]{Emanuel_1986}
Emanuel, K.~A., 1986: An air-sea interaction theory for tropical cyclones. {P}art {I}: {S}teady-state maintenance. \textit{Journal of the Atmospheric Sciences}, \textbf{43~(6)}, 585--605.

\bibitem[{Emanuel and Nolan(2004)Emanuel, and Nolan}]{Emanuel_Nolan_2004}
Emanuel, K.~A., and D.~S. Nolan, 2004: Tropical cyclone activity and the global climate system. \textit{26th Conf. on Hurricanes and Tropical Meteorology, American Meteor Society, 10A.2, Miami, FL}.

\bibitem[{Evans and Braun(2012)Evans, and Braun}]{Evans_Braun2012}
Evans, J.~L., and A.~Braun, 2012: A {Climatology} of {Subtropical} {Cyclones} in the {South} {Atlantic}. \textit{Journal of Climate}, \textbf{25~(21)}, 7328--7340, \doi{10.1175/JCLI-D-11-00212.1}.

\bibitem[{Eyring et~al.(2016)Eyring, Bony, Meehl, Senior, Stevens, Stouffer,, and Taylor}]{Eyring_etal_2016}
Eyring, V., S.~Bony, G.~A. Meehl, C.~A. Senior, B.~Stevens, R.~J. Stouffer, and K.~E. Taylor, 2016: {Overview of the Coupled Model Intercomparison Project Phase 6 (CMIP6) experimental design and organization}. \textit{Geoscientific Model Development}, \textbf{9~(5)}, 1937--1958, \doi{10.5194/gmd-9-1937-2016}, \urlprefix\url{https://gmd.copernicus.org/articles/9/1937/2016/}.

\bibitem[{Fu et~al.(2023)Fu, Chang,, and Liu}]{Fu_Chang_Liu2023}
Fu, D., P.~Chang, and X.~Liu, 2023: Using convolutional neural network to emulate seasonal tropical cyclone activity. \textit{Journal of Advances in Modeling Earth Systems}, \textbf{15}, e2022MS003\,596, \doi{10.1029/2022MS003596}.

\bibitem[{Garcia et~al.(2024)Garcia, Smith, Chavas,, and Komacek}]{Garcia_etal_2024}
Garcia, V., C.~Smith, D.~R. Chavas, and T.~D. Komacek, 2024: Tropical cyclones on tidally locked rocky planets: Dependence on rotation period. \textit{The Astrophysical Journal}, \textbf{Accepted}.

\bibitem[{Golaz et~al.(2019)}]{golazDOEE3SMCoupled2019}
Golaz, J.-C., and Coauthors, 2019: The {{DOE E3SM Coupled Model Version}} 1: {{Overview}} and {{Evaluation}} at {{Standard Resolution}}. \textit{J. Adv. Model. Earth Syst.}, \textbf{11~(7)}, 2089--2129, \doi{10.1029/2018MS001603}.

\bibitem[{Gozzo et~al.(2014)Gozzo, da~Rocha, Reboita,, and Sugahara}]{Gozzo_etal_2014}
Gozzo, L.~F., R.~P. da~Rocha, M.~S. Reboita, and S.~Sugahara, 2014: Subtropical {Cyclones} over the {Southwestern} {South} {Atlantic}: {Climatological} {Aspects} and {Case} {Study}. \textit{Journal of Climate}, \textbf{27~(22)}, 8543--8562, \doi{10.1175/JCLI-D-14-00149.1}.

\bibitem[{Gray(1979)}]{Gray_1979}
Gray, W.~M., 1979: Hurricanes: their formation, structure and likely role in the tropical circulation. meteorology over the tropical oceans. \textit{Roy. Meteor. Soc.}, 155--218.

\bibitem[{Gutjahr et~al.(2019)Gutjahr, Putrasahan, Lohmann, Jungclaus, {von Storch}, Br{\"u}ggemann, Haak,, and St{\"o}ssel}]{gutjahrMaxPlanckInstitute2019}
Gutjahr, O., D.~Putrasahan, K.~Lohmann, J.~H. Jungclaus, J.-S. {von Storch}, N.~Br{\"u}ggemann, H.~Haak, and A.~St{\"o}ssel, 2019: Max {{Planck Institute Earth System Model}} ({{MPI-ESM1}}.2) for the {{High-Resolution Model Intercomparison Project}} ({{HighResMIP}}). \textit{Geosci. Model. Dev.}, \textbf{12~(7)}, 3241--3281, \doi{10.5194/gmd-12-3241-2019}.

\bibitem[{Hajima et~al.(2020)}]{hajimaDevelopmentMIROCES2LEarth2020}
Hajima, T., and Coauthors, 2020: Development of the {{MIROC-ES2L Earth}} system model and the evaluation of biogeochemical processes and feedbacks. \textit{Geosci. Model. Dev.}, \textbf{13~(5)}, 2197--2244, \doi{10.5194/gmd-13-2197-2020}.

\bibitem[{He et~al.(2019)}]{heCASFGOALSf3LModel2019}
He, B., and Coauthors, 2019: {{CAS FGOALS-f3-L Model Datasets}} for {{CMIP6 Historical Atmospheric Model Intercomparison Project Simulation}}. \textit{Adv. Atmos. Sci.}, \textbf{36~(8)}, 771--778, \doi{10.1007/s00376-019-9027-8}.

\bibitem[{Held and Zhao(2008)Held, and Zhao}]{Held_Zhao_2008}
Held, I.~M., and M.~Zhao, 2008: Horizontally homogeneous rotating radiative--convective equilibria at {GCM} resolution. \textit{Journal of the Atmospheric Sciences}, \textbf{65~(6)}, 2003--2013.

\bibitem[{Held et~al.(2019)}]{heldStructurePerformanceGFDL2019}
Held, I.~M., and Coauthors, 2019: Structure and {{Performance}} of {{GFDL}}'s {{CM4}}.0 {{Climate Model}}. \textit{J. Adv. Model. Earth Syst.}, \textbf{11~(11)}, 3691--3727, \doi{10.1029/2019MS001829}.

\bibitem[{Hersbach et~al.(2020)}]{Hersbach_etal_2020}
Hersbach, H., and Coauthors, 2020: The {ERA5} global reanalysis. \textit{Quarterly Journal of the Royal Meteorological Society}, \textbf{146~(730)}, 1999--2049.

\bibitem[{Hoogewind et~al.(2019)Hoogewind, Chavas, Schenkel,, and O'Neill}]{Hoogewind_etal_2019}
Hoogewind, K., D.~R. Chavas, B.~Schenkel, and M.~O'Neill, 2019: Exploring environmental constraints on the observed global tropical cyclone count. \textit{Journal of Climate}, in review.

\bibitem[{Hsieh et~al.(2020)Hsieh, Vecchi, Yang, Held,, and Garner}]{Hsieh_eatl_2020}
Hsieh, T.-L., G.~A. Vecchi, W.~Yang, I.~M. Held, and S.~T. Garner, 2020: Large-scale control on the frequency of tropical cyclones and seeds: A consistent relationship across a hierarchy of global atmospheric models. \textit{Climate Dynamics}, \textbf{55~(11-12)}, 3177--3196.

\bibitem[{Kelley et~al.(2020)}]{kelleyGISSE2ConfigurationsClimatology2020}
Kelley, M., and Coauthors, 2020: {{GISS-E2}}.1: {{Configurations}} and {{Climatology}}. \textit{J. Adv. Model. Earth Syst.}, \textbf{12~(8)}, e2019MS002\,025, \doi{10.1029/2019MS002025}.

\bibitem[{Khairoutdinov and Emanuel(2013)Khairoutdinov, and Emanuel}]{Khairoutdinov_Emanuel_2013}
Khairoutdinov, M., and K.~Emanuel, 2013: Rotating radiative-convective equilibrium simulated by a cloud-resolving model. \textit{Journal of Advances in Modeling Earth Systems}, \textbf{5~(4)}, 816--825.

\bibitem[{Klotzbach et~al.(2020)Klotzbach, Bell, Bowen, Gibney, Knapp,, and Schreck~III}]{Klotzbach_etal_2020}
Klotzbach, P.~J., M.~M. Bell, S.~G. Bowen, E.~J. Gibney, K.~R. Knapp, and C.~J. Schreck~III, 2020: Surface pressure a more skillful predictor of normalized hurricane damage than maximum sustained wind. \textit{Bulletin of the American Meteorological Society}, \textbf{101~(6)}, E830--E846.

\bibitem[{Klotzbach and Oliver(2015)Klotzbach, and Oliver}]{Klotzbach_Oliver_2015}
Klotzbach, P.~J., and E.~C. Oliver, 2015: Modulation of atlantic basin tropical cyclone activity by the madden--julian oscillation (mjo) from 1905 to 2011. \textit{Journal of Climate}, \textbf{28~(1)}, 204--217.

\bibitem[{Knapp et~al.(2010)Knapp, Kruk, Levinson, Diamond,, and Neumann}]{Knapp_etal_2010}
Knapp, K.~R., M.~C. Kruk, D.~H. Levinson, H.~J. Diamond, and C.~J. Neumann, 2010: The international best track archive for climate stewardship (ibtracs) unifying tropical cyclone data. \textit{Bulletin of the American Meteorological Society}, \textbf{91~(3)}, 363--376.

\bibitem[{Knutson et~al.(2020)}]{Knutson_etal_2020}
Knutson, T., and Coauthors, 2020: Tropical cyclones and climate change assessment: Part {II}: Projected response to anthropogenic warming. \textit{Bulletin of the American Meteorological Society}, \textbf{101~(3)}, E303--E322.

\bibitem[{Komacek et~al.(2020)Komacek, Chavas,, and Abbot}]{Komacek_Chavas_Abbot_2020}
Komacek, T.~D., D.~R. Chavas, and D.~S. Abbot, 2020: Hurricane genesis is favorable on terrestrial exoplanets orbiting late-type m dwarf stars. \textit{The Astrophysical Journal}, \textbf{898~(2)}, 115.

\bibitem[{Korty et~al.(2012)Korty, Camargo,, and Galewsky}]{Korty_Camargo_Galewsky_2012}
Korty, R.~L., S.~J. Camargo, and J.~Galewsky, 2012: Tropical cyclone genesis factors in simulations of the last glacial maximum. \textit{Journal of Climate}, \textbf{25~(12)}, 4348--4365.

\bibitem[{Korty et~al.(2017)Korty, Emanuel, Huber,, and Zamora}]{Korty_etal_2017}
Korty, R.~L., K.~A. Emanuel, M.~Huber, and R.~A. Zamora, 2017: Tropical cyclones downscaled from simulations with very high carbon dioxide levels. \textit{Journal of Climate}, \textbf{30~(2)}, 649--667.

\bibitem[{Kossin et~al.(2016)Kossin, Emanuel,, and Camargo}]{Kossin_etal_2016}
Kossin, J.~P., K.~A. Emanuel, and S.~J. Camargo, 2016: Past and {Projected} {Changes} in {Western} {North} {Pacific} {Tropical} {Cyclone} {Exposure}. \textit{Journal of Climate}, \textbf{29~(16)}, 5725--5739, \doi{10.1175/JCLI-D-16-0076.1}.

\bibitem[{Kossin et~al.(2014)Kossin, Emanuel,, and Vecchi}]{Kossin_Emanuel_Vecchi_2014}
Kossin, J.~P., K.~A. Emanuel, and G.~A. Vecchi, 2014: The poleward migration of the location of tropical cyclone maximum intensity. \textit{Nature}, \textbf{509~(7500)}, 349--352.

\bibitem[{Kuhlbrodt et~al.(2018)}]{kuhlbrodtLowResolutionVersionHadGEM32018}
Kuhlbrodt, T., and Coauthors, 2018: The {{Low-Resolution Version}} of {{HadGEM3 GC3}}.1: {{Development}} and {{Evaluation}} for {{Global Climate}}. \textit{J. Adv. Model. Earth Syst.}, \textbf{10~(11)}, 2865--2888, \doi{10.1029/2018MS001370}.

\bibitem[{Lee et~al.(2020{\natexlab{a}})Lee, Camargo, Sobel,, and Tippett}]{Lee_etal_2020}
Lee, C.-Y., S.~J. Camargo, A.~H. Sobel, and M.~K. Tippett, 2020{\natexlab{a}}: Statistical--dynamical downscaling projections of tropical cyclone activity in a warming climate: Two diverging genesis scenarios. \textit{Journal of Climate}, \textbf{33~(11)}, 4815--4834.

\bibitem[{Lee et~al.(2022)Lee, Sobel, Camargo, Tippett,, and Yang}]{Lee_etal_2022}
Lee, C.-Y., A.~H. Sobel, S.~J. Camargo, M.~K. Tippett, and Q.~Yang, 2022: New {{York State Hurricane Hazard}}: {{History}} and {{Future Projections}}. \textit{Journal of Applied Meteorology and Climatology}, \textbf{61~(6)}, 613--629, \doi{10.1175/JAMC-D-21-0173.1}.

\bibitem[{Lee et~al.(2023)Lee, Sobel, Tippett, Camargo, W\"uest, Wehner,, and Murakami}]{Lee_etal_2023}
Lee, C.-Y., A.~H. Sobel, M.~K. Tippett, S.~J. Camargo, M.~W\"uest, M.~Wehner, and H.~Murakami, 2023: Climate change signal in {Atlantic} tropical cyclones today and near future. \textit{Earth's Future}, \textbf{11}, e2023EF003\,539, \doi{10.1029/2023EF003539}.

\bibitem[{Lee et~al.(2020{\natexlab{b}})Lee, Kim, Sun, Kim, Moon, Sung, Kim,, and Byun}]{leeEvaluationKoreaMeteorological2020}
Lee, J., J.~Kim, M.-A. Sun, B.-H. Kim, H.~Moon, H.~M. Sung, J.~Kim, and Y.-H. Byun, 2020{\natexlab{b}}: Evaluation of the {{Korea Meteorological Administration Advanced Community Earth-System}} model ({{K-ACE}}). \textit{Asia-Pacific J. Atmos. Sci.}, \textbf{56~(3)}, 381--395, \doi{10.1007/s13143-019-00144-7}.

\bibitem[{Li et~al.(2020)}]{liFlexibleGlobalOceanAtmosphereLand2020}
Li, L., and Coauthors, 2020: The {{Flexible Global Ocean-Atmosphere-Land System Model Grid-Point Version}} 3 ({{FGOALS-g3}}): {{Description}} and {{Evaluation}}. \textit{J. Adv. Model. Earth Syst.}, \textbf{12~(9)}, e2019MS002\,012, \doi{10.1029/2019MS002012}.

\bibitem[{Lin et~al.(2023)Lin, Camargo, Lien, Shi,, and Kossin}]{Lin_etal_2023}
Lin, I.-I., S.~J. Camargo, C.-C. Lien, C.-A. Shi, and J.~P. Kossin, 2023: Poleward migration as global warming’s possible self-regulator to restrain future western {North} {Pacific} {Tropical} {Cyclone}’s intensification. \textit{npj Climate and Atmospheric Science}, \textbf{6~(1)}, 1--9, \doi{10.1038/s41612-023-00329-y}.

\bibitem[{Lin et~al.(2020)}]{linCommunityIntegratedEarth2020}
Lin, Y., and Coauthors, 2020: Community {{Integrated Earth System Model}} ({{CIESM}}): {{Description}} and {{Evaluation}}. \textit{J. Adv. Model. Earth Syst.}, \textbf{12~(8)}, e2019MS002\,036, \doi{10.1029/2019MS002036}.

\bibitem[{Lovato et~al.(2022)}]{lovatoCMIP6SimulationsCMCC2022}
Lovato, T., and Coauthors, 2022: {{CMIP6 Simulations With}} the {{CMCC Earth System Model}} ({{CMCC-ESM2}}). \textit{J. Adv. Model. Earth Syst.}, \textbf{14~(3)}, e2021MS002\,814, \doi{10.1029/2021MS002814}.

\bibitem[{Lu et~al.(2021)Lu, Ge,, and Peng}]{Lu_Ge_Peng_2021}
Lu, C., X.~Ge, and M.~Peng, 2021: Comparison of controlling parameters for near-equatorial tropical cyclone formation between western north pacific and north atlantic. \textit{Journal of Meteorological Research}, \textbf{35~(4)}, 623--634.

\bibitem[{Mauritsen et~al.(2019)}]{mauritsenDevelopmentsMPIMEarth2019a}
Mauritsen, T., and Coauthors, 2019: Developments in the {{MPI-M Earth System Model}} version 1.2 ({{MPI-ESM1}}.2) and {{Its Response}} to {{Increasing CO2}}. \textit{J. Adv. Model. Earth Syst.}, \textbf{11~(4)}, 998--1038, \doi{10.1029/2018MS001400}.

\bibitem[{Mei et~al.(2019)Mei, Kamae, Xie,, and Yoshida}]{Mei_etal_2019}
Mei, W., Y.~Kamae, S.-P. Xie, and K.~Yoshida, 2019: Variability and predictability of north atlantic hurricane frequency in a large ensemble of high-resolution atmospheric simulations. \textit{Journal of Climate}, \textbf{32~(11)}, 3153--3167.

\bibitem[{Meng and Garner(2023)Meng, and Garner}]{Meng_Garner_2023}
Meng, L., and S.~T. Garner, 2023: Nonlocal controls on tropical cyclogenesis: A trajectory-based genesis potential index. \textit{Journal of the Atmospheric Sciences}, \textbf{80}, 2925 -- 2946, \doi{10.1175/JAS-D-23-0025.1}.

\bibitem[{Merlis et~al.(2013)Merlis, Zhao,, and Held}]{Merlis_Zhao_Held_2013}
Merlis, T.~M., M.~Zhao, and I.~M. Held, 2013: The sensitivity of hurricane frequency to itcz changes and radiatively forced warming in aquaplanet simulations. \textit{Geophysical Research Letters}, \textbf{40~(15)}, 4109--4114.

\bibitem[{Miller et~al.(2021)}]{millerCMIP6HistoricalSimulations2021}
Miller, R.~L., and Coauthors, 2021: {{CMIP6 Historical Simulations}} (1850–2014) {{With GISS-E2}}.1. \textit{J. Adv. Model. Earth Syst.}, \textbf{13~(1)}, e2019MS002\,034, \doi{10.1029/2019MS002034}.

\bibitem[{Müller et~al.(2018)}]{mullerHigherresolutionVersionMax2018}
Müller, W.~A., and Coauthors, 2018: A {{Higher-resolution Version}} of the {{Max Planck Institute Earth System Model}} ({{MPI-ESM1}}.2-{{HR}}). \textit{J. Adv. Model. Earth Syst.}, \textbf{10~(7)}, 1383--1413, \doi{10.1029/2017MS001217}.

\bibitem[{Pak et~al.(2021)}]{pakKoreaInstituteOcean2021}
Pak, G., and Coauthors, 2021: Korea {{Institute}} of {{Ocean Science}} and {{Technology Earth System Model}} and {{Its Simulation Characteristics}}. \textit{Ocean Sci. J.}, \textbf{56~(1)}, 18--45, \doi{10.1007/s12601-021-00001-7}.

\bibitem[{Patricola et~al.(2014)Patricola, Saravanan,, and Chang}]{Patricola_Saravanan_Chang_2014}
Patricola, C.~M., R.~Saravanan, and P.~Chang, 2014: The impact of the el ni{\~n}o--southern oscillation and {Atlantic Meridional Mode} on seasonal {Atlantic} tropical cyclone activity. \textit{Journal of Climate}, \textbf{27~(14)}, 5311--5328.

\bibitem[{Pausata and Camargo(2019)Pausata, and Camargo}]{Pausata_Camargo_2019}
Pausata, F.~S., and S.~J. Camargo, 2019: Tropical cyclone activity affected by volcanically induced itcz shifts. \textit{Proceedings of the National Academy of Sciences}, \textbf{116~(16)}, 7732--7737.

\bibitem[{Pausata et~al.(2017)Pausata, Emanuel, Chiacchio, Diro, Zhang, Sushama, Stager,, and Donnelly}]{Pausata_etal_2017}
Pausata, F.~S., K.~A. Emanuel, M.~Chiacchio, G.~T. Diro, Q.~Zhang, L.~Sushama, J.~C. Stager, and J.~P. Donnelly, 2017: Tropical cyclone activity enhanced by sahara greening and reduced dust emissions during the african humid period. \textit{Proceedings of the National Academy of Sciences}, \textbf{114~(24)}, 6221--6226.

\bibitem[{Pezza and Simmonds(2005)Pezza, and Simmonds}]{Pezza_Simmonds_2005}
Pezza, A.~B., and I.~Simmonds, 2005: The first south atlantic hurricane: Unprecedented blocking, low shear and climate change. \textit{Geophysical Research Letters}, \textbf{32~(15)}.

\bibitem[{Pielke~Jr et~al.(2022)Pielke~Jr, Burgess,, and Ritchie}]{Pielke_Burgess_Ritchie_2022}
Pielke~Jr, R., M.~G. Burgess, and J.~Ritchie, 2022: Plausible 2005-2050 emissions scenarios project between 2 and 3 degrees c of warming by 2100. \textit{Environmental Research Letters}.

\bibitem[{Rong et~al.(2021)Rong, Li, Chen, Su, Hua, Zhang,, and Xin}]{rongCMIP6HistoricalSimulation2021}
Rong, X., J.~Li, H.~Chen, J.~Su, L.~Hua, Z.~Zhang, and Y.~Xin, 2021: The {{CMIP6 Historical Simulation Datasets Produced}} by the {{Climate System Model CAMS-CSM}}. \textit{Adv. Atmos. Sci.}, \textbf{38~(2)}, 285--295, \doi{10.1007/s00376-020-0171-y}.

\bibitem[{Rong et~al.(2018)}]{rongCAMSClimateSystem2018}
Rong, X., and Coauthors, 2018: The {{CAMS Climate System Model}} and a {{Basic Evaluation}} of {{Its Climatology}} and {{Climate Variability Simulation}}. \textit{J. Meteorol. Res.}, \textbf{32~(6)}, 839--861, \doi{10.1007/s13351-018-8058-x}.

\bibitem[{Rousseau-Rizzi and Emanuel(2019)Rousseau-Rizzi, and Emanuel}]{Rousseaurizzi_Emanuel_2019}
Rousseau-Rizzi, R., and K.~Emanuel, 2019: An evaluation of hurricane superintensity in axisymmetric numerical models. \textit{Journal of the Atmospheric Sciences}, \textbf{76~(6)}, 1697--1708.

\bibitem[{Seager et~al.(2019)Seager, Cane, Henderson, Lee, Abernathey,, and Zhang}]{Seager_etal_2019}
Seager, R., M.~Cane, N.~Henderson, D.-E. Lee, R.~Abernathey, and H.~Zhang, 2019: Strengthening tropical {Pacific} zonal sea surface temperature gradient consistent with rising greenhouse gases. \textit{Nature Clim.ate Change}, \textbf{9~(7)}, 517--522, \doi{10.1038/s41558-019-0505-x}.

\bibitem[{Seland et~al.(2020)}]{selandOverviewNorwegianEarth2020}
Seland, {\O}., and Coauthors, 2020: Overview of the {{Norwegian Earth System Model}} ({{NorESM2}}) and key climate response of {{CMIP6 DECK}}, historical, and scenario simulations. \textit{Geosci. Model. Dev.}, \textbf{13~(12)}, 6165--6200, \doi{10.5194/gmd-13-6165-2020}.

\bibitem[{Sellar et~al.(2019)}]{sellarUKESM1DescriptionEvaluation2019}
Sellar, A.~A., and Coauthors, 2019: {{UKESM1}}: {{Description}} and {{Evaluation}} of the {{U}}.{{K}}. {{Earth System Model}}. \textit{J. Adv. Model. Earth Syst.}, \textbf{11~(12)}, 4513--4558, \doi{10.1029/2019MS001739}.

\bibitem[{Semmler et~al.(2020)}]{semmlerSimulationsCMIP6AWI2020}
Semmler, T., and Coauthors, 2020: Simulations for {{CMIP6 With}} the {{AWI Climate Model AWI-CM-1-1}}. \textit{J. Adv. Model. Earth Syst.}, \textbf{12~(9)}, e2019MS002\,009, \doi{10.1029/2019MS002009}.

\bibitem[{Sobel et~al.(2021)Sobel, Wing, Camargo, Patricola, Vecchi, Lee,, and Tippett}]{Sobel_etal_2021}
Sobel, A.~H., A.~A. Wing, S.~J. Camargo, C.~M. Patricola, G.~A. Vecchi, C.-Y. Lee, and M.~K. Tippett, 2021: Tropical cyclone frequency. \textit{Earth's Future}, \textbf{9~(12)}, e2021EF002\,275.

\bibitem[{Sobel et~al.(2023)}]{Sobel_etal_2023}
Sobel, A.~H., and Coauthors, 2023: Near-term tropical cyclone risk and coupled earth system model biases. \textit{Proceedings of the National Academy of Sciences}, \textbf{120~(33)}, e2209631\,120.

\bibitem[{Studholme et~al.(2022)Studholme, Fedorov, Gulev, Emanuel,, and Hodges}]{Studholme_etal_2022}
Studholme, J., A.~V. Fedorov, S.~K. Gulev, K.~Emanuel, and K.~Hodges, 2022: Poleward expansion of tropical cyclone latitudes in warming climates. \textit{Nature Geoscience}, \textbf{15~(1)}, 14--28.

\bibitem[{Sugi et~al.(2020)Sugi, Yamada, Yoshida, Mizuta, Nakano, Kodama,, and Satoh}]{Sugietal2020}
Sugi, M., Y.~Yamada, K.~Yoshida, R.~Mizuta, M.~Nakano, C.~Kodama, and M.~Satoh, 2020: Future changes in the global frequency of tropical cyclone seeds. \textit{Sola}, \textbf{16}, 70--74.

\bibitem[{Swart et~al.(2019)}]{swartCanadianEarthSystem2019}
Swart, N.~C., and Coauthors, 2019: The {{Canadian Earth System Model}} version 5 ({{CanESM5}}.0.3). \textit{Geosci. Model. Dev.}, \textbf{12~(11)}, 4823--4873, \doi{10.5194/gmd-12-4823-2019}.

\bibitem[{Séférian et~al.(2019)}]{seferianEvaluationCNRMEarth2019}
Séférian, R., and Coauthors, 2019: Evaluation of {{CNRM Earth System Model}}, {{CNRM-ESM2-1}}: {{Role}} of {{Earth System Processes}} in {{Present-Day}} and {{Future Climate}}. \textit{J. Adv. Model. Earth Syst.}, \textbf{11~(12)}, 4182--4227, \doi{10.1029/2019MS001791}.

\bibitem[{Tang and Emanuel(2010)Tang, and Emanuel}]{Tang_Emanuel_2010}
Tang, B., and K.~Emanuel, 2010: Midlevel ventilation's constraint on tropical cyclone intensity. \textit{Journal of the Atmospheric Sciences}, \textbf{67~(6)}, 1817--1830.

\bibitem[{Tang and Emanuel(2012)Tang, and Emanuel}]{Tang_Emanuel_2012}
Tang, B., and K.~Emanuel, 2012: A ventilation index for tropical cyclones. \textit{Bulletin of the American Meteorological Society}, \textbf{93~(12)}, 1901--1912.

\bibitem[{Tatebe et~al.(2019)}]{tatebeDescriptionBasicEvaluation2019}
Tatebe, H., and Coauthors, 2019: Description and basic evaluation of simulated mean state, internal variability, and climate sensitivity in {{MIROC6}}. \textit{Geosci. Model. Dev.}, \textbf{12~(7)}, 2727--2765, \doi{10.5194/gmd-12-2727-2019}.

\bibitem[{Tippett et~al.(2011)Tippett, Camargo,, and Sobel}]{Tippett_Camargo_Sobel_2011}
Tippett, M.~K., S.~J. Camargo, and A.~H. Sobel, 2011: A poisson regression index for tropical cyclone genesis and the role of large-scale vorticity in genesis. \textit{Journal of Climate}, \textbf{24~(9)}, 2335--2357.

\bibitem[{Tory et~al.(2018)Tory, Ye,, and Dare}]{Tory_Ye_Dare_2018}
Tory, K.~J., H.~Ye, and R.~Dare, 2018: Understanding the geographic distribution of tropical cyclone formation for applications in climate models. \textit{Climate Dynamics}, \textbf{50}, 2489--2512.

\bibitem[{Vecchi et~al.(2021)Vecchi, Landsea, Zhang, Villarini,, and Knutson}]{Vecchi_etal_2021}
Vecchi, G.~A., C.~Landsea, W.~Zhang, G.~Villarini, and T.~Knutson, 2021: Changes in {Atlantic} major hurricane frequency since the late-19th century. \textit{Nature Communications}, \textbf{12~(1)}, 4054, \doi{10.1038/s41467-021-24268-5}.

\bibitem[{Voldoire et~al.(2019)}]{voldoireEvaluationCMIP6DECK2019}
Voldoire, A., and Coauthors, 2019: Evaluation of {{CMIP6 DECK Experiments With CNRM-CM6-1}}. \textit{J. Adv. Model. Earth Syst.}, \textbf{11~(7)}, 2177--2213, \doi{10.1029/2019MS001683}.

\bibitem[{Volodin et~al.(2010)Volodin, Dianskii,, and Gusev}]{volodinSimulatingPresentdayClimate2010a}
Volodin, E.~M., N.~A. Dianskii, and A.~V. Gusev, 2010: Simulating present-day climate with the {{INMCM4}}.0 coupled model of the atmospheric and oceanic general circulations. \textit{Izv. Atmos. Ocean. Phys.}, \textbf{46~(4)}, 414--431, \doi{10.1134/S000143381004002X}.

\bibitem[{Volodin et~al.(2017)}]{volodinSimulationPresentdayClimate2017}
Volodin, E.~M., and Coauthors, 2017: Simulation of the present-day climate with the climate model {{INMCM5}}. \textit{Clim. Dyn.}, \textbf{49~(11)}, 3715--3734, \doi{10.1007/s00382-017-3539-7}.

\bibitem[{Wang and Moon(2017)Wang, and Moon}]{Wang_Moon_2017}
Wang, B., and J.-Y. Moon, 2017: An anomalous genesis potential index for mjo modulation of tropical cyclones. \textit{Journal of Climate}, \textbf{30~(11)}, 4021--4035.

\bibitem[{Wang and Murakami(2020)Wang, and Murakami}]{Wang_Murakami_2020}
Wang, B., and H.~Murakami, 2020: Dynamical genesis potential index for diagnosing present-day and future global tropical cyclone genesis. \textit{Environmental Research Letters}, \textbf{15}, 114\,008, \doi{10.1088/1748-9326/1bbb01}.

\bibitem[{Wang et~al.(2021)}]{wangPerformanceTaiwanEarth2021}
Wang, Y.-C., and Coauthors, 2021: Performance of the {{Taiwan Earth System Model}} in {{Simulating Climate Variability Compared With Observations}} and {{CMIP6 Model Simulations}}. \textit{J. Adv. Model. Earth Syst.}, \textbf{13~(7)}, e2020MS002\,353, \doi{10.1029/2020MS002353}.

\bibitem[{Wu et~al.(2019)}]{wuBeijingClimateCenter2019}
Wu, T., and Coauthors, 2019: The {{Beijing Climate Center Climate System Model}} ({{BCC-CSM}}): The main progress from {{CMIP5}} to {{CMIP6}}. \textit{Geosci. Model. Dev.}, \textbf{12~(4)}, 1573--1600, \doi{10.5194/gmd-12-1573-2019}.

\bibitem[{Yukimoto et~al.(2019)}]{yukimotoMeteorologicalResearchInstitute2019}
Yukimoto, S., and Coauthors, 2019: The {{Meteorological Research Institute Earth System Model Version}} 2.0, {{MRI-ESM2}}.0: {{Description}} and {{Basic Evaluation}} of the {{Physical Component}}. \textit{J. Meteorol. Soc. Japan.}, \textbf{97~(5)}, 931--965, \doi{10.2151/jmsj.2019-051}.

\bibitem[{Zhang et~al.(2021)Zhang, Soden, Vecchi,, and Yang}]{Zhang_etal_2021b}
Zhang, B., B.~J. Soden, G.~A. Vecchi, and W.~Yang, 2021: The role of radiative interactions in tropical cyclone development under realistic boundary conditions. \textit{Journal of climate}, \textbf{34~(6)}, 2079--2091.

\bibitem[{Zhang et~al.(2020)}]{zhangDescriptionClimateSimulation2020}
Zhang, H., and Coauthors, 2020: Description and {{Climate Simulation Performance}} of {{CAS-ESM Version}} 2. \textit{J. Adv. Model. Earth Syst.}, \textbf{12~(12)}, e2020MS002\,210, \doi{10.1029/2020MS002210}.

\bibitem[{Ziehn et~al.(2020)}]{ziehnAustralianEarthSystem2020}
Ziehn, T., and Coauthors, 2020: The {{Australian Earth System Model}}: {{ACCESS-ESM1}}.5. \textit{J. South. Hemisphere Earth Syst. Sci.}, \textbf{70}, 193--214, \doi{10.1071/ES19035}.

\end{thebibliography}

\end{document}